\documentstyle[12pt,epsfig,twoside]{article}

\begin{document}
\noindent
{\Large\bf
Chiral symmetry breaking in  hot matter.}\footnote{Lectures given at
the eleventh Chris Engelbrecht Summer School in Theoretical Physics,
4-13 February, 1998.}

\vskip 0.4in
\noindent
S.P. Klevansky\footnote{internet
address:{\tt sandi@frodo.tphys.uni-heidelberg.de}}

\vskip 0.2in
\noindent
Institut f\"ur Theoretische Physik, Philosophenweg 19, D-69120 Heidelberg,
Germany.

\vskip 0.4in

\noindent
{\bf Abstract}  
This series of three lectures covers (a) a basic introduction to 
symmetry breaking in general and chiral symmetry breaking in QCD,   (b) an
overview of the present status of lattice data and the knowledge
that we have at finite temperature from chiral perturbation theory.
(c) Results obtained from the Nambu--Jona-Lasinio model describing
static mesonic properties are discussed as well as the bulk thermodynamic
quantities.    Divergences that are observed in the elastic
quark-antiquark scattering cross-section, reminiscent of the phenomenon
of critical opalescence in light scattering, is also discussed.
(d) Finally, we deal with the realm of systems out of equilibrium,
and examine the effects of a medium dependent condensate in  
a system of interacting quarks.  

\vfill

\eject
\tableofcontents

\eject  

\section{Introduction}\label{secintro}

Chiral symmetry is the symmetry of Quantum Chromodynamics (QCD) that
dictates the static properties of the low lying mesonic sector, in
particular those pertaining to the pseudoscalar nonet $(\pi,K,\eta)$.
This symmetry is responsible for the fact that, in its broken phase,
quarks acquire mass (and are termed ``constituent'' quarks, as they
form parts of hadrons, while, in the restored phase, quarks have only
their small or current mass values.
It is believed that at finite temperature this symmetry is restored, a
feature that is strongly motivated by numerical studies of QCD on the 
lattice. 
Concomitantly with this picture, it is believed that another phase transition
from a deconfined phase of matter (consisting of a
 hot fireball of quarks and gluons)
to a confined phase can occur, in which only the final state of hadrons
is observed. 
  Given  these two features, a large amount of scientific endeavor has
been and will continue to be invested in the study of heavy ion collisions,
in which high temperatures can be attained.   In particular, the low-lying
mesons are copiously produced, and since these provide the testing ground
for chiral symmetry at $T=0$, it is hoped that (with enough theoretical
and experimental study), a clear signal of this phase transition will
emerge.  To be quite precise, one requires unambiguous signals of
{\it both} phase transitions, that of confinement/deconfinement, as well
as chiral symmetry breaking/restoration.
 Thus far, however, there are no  unambiguous signals known
that are experimentally measurable for either of these transitions.
In this paper, we shall confine ourselves to a discussion of chiral symmetry
and its associated aspects, leaving the difficulties of confinement to a
later stage.    

This series of three  lectures is intended to introduce the concepts of
chiral symmetry starting  from basics.   There is a short
  guide for  the uninitiated into the ideas
of
what symmetry breaking 
 is, and then an attempt to  summarize the current status of what we know to
be fact, taken on the theoretical level, at finite temperature.
This involves examining firstly the lattice gauge simulations of QCD at finite
temperature and then examining how far we can go with chiral perturbation 
theory \cite{chpt,gasser}.  
From lattice gauge simulations, the existence of the chiral and 
deconfinement phase transitions is inferred.   Critical exponents for the
chiral
transition have been obtained, but are as yet not conclusive.   Temperature
dependence of the mesonic screening masses have also been calculated, and
the question of $U_A(1)$ symmetry restoration addressed.   Bulk
thermodynamic properties have been studied over several years, with larger
and larger lattices, and this represents the state of the art of what
we know today about these quantities in QCD.   By contrast, while chiral
perturbation theory gives a superb description of the low energy sector,
and also gives the leading behavior expected of the order parameter as a 
function of temperature \cite{gerber,toublan}, it cannot {\it per se} be
used to describe the phase transition region, which is non-analytic.
The level of accuracy of CHPT at finite temperatures is  illustrated in 
the calculation of the pion masses as a function of temperature in a 
recent publication \cite{toublan}.

 Note that we  restrict ourselves mainly to 
finite temperature and not  to finite density.
   The first lattice simulations at
finite density have already been performed \cite{barbour}.    However, there are 
many technical difficulties that are not yet under control, and as such
no results are completely reliable as yet.     For this reason, we will
also not attempt to make any model discussions at finite density at this
stage,
although there are of course several.

In the second lecture,    a simple chiral model, the Nambu--Jona-Lasinio
(NJL) \cite{njl,spk,reviews} model is discussed,
in which it becomes evident that features relating to static properties of
the low-lying mesons are excellently reproduced.
This includes charge radii, meson-meson scattering lengths,
polarizabilities,
{\it etc}, and one can validate that the expected results of chiral
perturbation theory are recovered, here with very few parameters.  In
addition, the variation of the meson masses with temperature, although
calculated in this model
 as {\it pole} masses, shows the same qualitative behavior 
as was observed by the lattice gauge groups.    Given these successes 
    with this model, one is encouraged to
study the dependence of all static mesonic 
properties as a function of the
temperature in order 
to investigate whether abrupt behavior occurs at the phase
transition point.
For two flavors of quarks, one finds that the pseudoscalar sector in
particular is typified by an almost constant behavior in all static
properties (such as the mean pion radius $\langle r_\pi^2\rangle^{1/2}$,
the polarizabilities $\alpha_\pi$ and the scattering lengths $a_\pi$)
for a wide range of the temperature shortly up until the point
at which the chiral
 phase transition occurs, and then these quantities show a sharp divergence. 
This is true for the case in which the current quark mass
of the up and down quarks $m_0^u=m_0^d=m_0=0$, and a
phase transition can occur.   When $m_0\ne 0$, only a crossover can be 
observed in the order parameter.    A new
 transition temperature $T_M=T_{{\rm Mott}}$
is defined as being the temperature at which the mesonic states become
unbound, or resonances.   It thus respresents a delocalization of the
mesons, rather than their deconfinement.   The static properties of the 
pionic sector then remain constant for most of the temperature range, and
diverge at the Mott temperature.  One thus still observes  
 a dramatic structure -- 
either directly at the phase transition temperature itself
in the case of $m_0=0$ or at $T=T_M$ for $m_0\ne0$.

It is also extremely  interesting to study 
 dynamical quantities such as the elastic cross-section for
$q\bar q\rightarrow q\bar q$  scattering.    This particular quantity  
displays a divergence at the critical or Mott temperature  in a similar
fashion  as occurs in the
phenomenon of critical opalescence that is observed in the scattering of
light.  However, although this feature and those observed for the static
properties
 are exciting, their direct measurement is
elusive if not downright impossible.   

The scalar mesonic sector within the NJL model is observed to display
 a completely different behavior.  Here
the mass drops relatively quickly with temperature.    Nevertheless, 
experimentally, the scalar mesons constitute a multiplet that appears to
have the symmetry badly broken, and the lowest meson of which (the $\sigma$)
has an extremely large width.    Consequently only indirect information on
this sector is useful.

How then can one hope to observe the chiral phase transition?     To attempt
to answer this question, we recall that the chiral phase transition appears
to be intimately linked with the confinement/deconfinement phase transition,
i.e. they appear to take place at the same temperature \cite{hwa}.
A heuristic understanding of this feature is quite satisfactory -- it 
implies that at high temperatures,
one should have chiral symmetry restored in a plasma phase, with free
(current) quarks and gluons being the ingredients, while at $T<T_c$, the
confined phase contains only hadrons that are made up of constituent 
(massive) quarks.    Experimental effort to detect the quark-gluon plasma 
phase is concentrated on contructing hot and dense matter via heavy-ion
collisions such as $Pb+Pb$ at increasingly high energies, and will form
a main part of the program of the two accelerators RHIC at Brookhaven and
the LHC (Geneva) that are currently under construction.
Given the fact now  that
heavy-ion collisions take place over a small time scale, it is conceivable
that the features of divergences occurring in both static and dynamical
quantities
 might enter realistically into a {\it non-equilibrium} treatment
of such collisions, which of course involves
many particles, the lightest of which are the pions,
 and thus to measurable observables.

For this reason, the final lecture is devoted to a discussion of 
 non-equi\-li\-bri\-um physics of an
interacting fermionic Lagrangian, and which is then applied to the
Nambu--Jona-Lasinio model in the lowest possible terms in an appropriate
double expansion in both $\hbar$ and the inverse number of colors $N_c$
\cite{quack,lemmer}.
Using the simplest approximations that lead
to a semi-classical result, one can recover a 
 Boltzmann like equation for the quark
distribution function.   Here one sees that the problems are simply open
ended.   The issue of constructing interlinked equations dealing with
several species of particle must be confronted and 
 the issue of multiparticle production
(hadronization) must be addressed,  since the usual Boltzman
 collision scenario that incorporates only binary collisions 
 is inadequate for a relativistic description.

Obviously it is an impossible task to  discuss all aspects of 
chiral symmetry breaking and restoration within three lectures, and for this
reason I have been highly selective in the material presented.   There are
many, many studies in the literature involving chiral symmetry, and
I am in no way attempting in this paper to be comprehensive.    The 
interested reader may also refer to the work of Refs.\cite{schaefer}
 for treatments
of the linear sigma model at finite temperature, for example,
 and to the work of Ref.
\cite{reinhardt}
for discussions in the baryonic sector, in addition to the other
general references that are given in the text.

The structure of this manuscript reflects the three lectures directly:  in 
Section 2, current factual information on the chiral transition, taken from
lattice gauge simulations and chiral perturbation theory is presented.
In Section 3, the Nambu--Jona-Lasinio (NJL) model is used to present
the ramifications of symmetry breaking at the critical temperature.   In 
Section 4, a non-equilibrium formulation of a theory of interacting fermions
is described and the equations are investigated for the NJL model.
In the concluding section, we discuss  where this could possibly lead to 
observable consequences.

\section{Equilibrium thermodynamics.}

In this section, we attempt to present those aspects of chiral symmetry
at finite temperature that are regarded as being ``exact'' or
factual, that is to say, they are derived from QCD itself, or from
considerations thereof.
   We start by briefly introducing the reader to the general
concept of symmetry breaking at $T=0$.   Following this, chiral 
symmetry breaking in the QCD Lagrangian is analysed.    In the 
following subsection, the simulations of lattice gauge theory are
discussed, dealing firstly with the temperature dependence of the order
parameter, the critical exponents
 obtained at the phase transition, meson screening masses and
the question of whether
$U_A(1)$ symmetry is restored at high temperatures or not.
Secondly, we indicate what is known from the lattice about
 bulk thermodynamic properties.   The pressure density, energy density
and entropy densities have been calculated on the lattice.   These
quantities give rather indications of the confinement/deconfinement 
transition, and as we will show in Section~3, cannot be described well
by a model that contains chiral symmetry alone, and which ignores the
confinement aspect.

   In the
final subsection, we briefly introduce the concepts of chiral 
perturbation theory (CHPT) and we describe  
the state of the art results at finite
temperature.  As will be seen, these give an important functional dependence
at low temperatures, but cannot be expected to cope with the phase
transition region, which is non-analytic.

\subsection{Introduction to chiral symmetry at $T=0$.}

The fact that a Hamiltonian, or equivalently a Lagrangian, is invariant
under a symmetry transformation results in a degeneracy within the spectrum
that is observed.     Mathematically, one expresses the fact that  a 
Hamilton function H is invariant under a specific symmetry
via the statement
\begin{equation}
UHU^\dagger = H
\label{e:usym}
\end{equation}
where $U$ is an
element of the group corresponding to this symmetry.   Now if one considers
the states $|A\rangle$ and $|B\rangle$ that are related by the
transformation $U$,
\begin{equation}
|B\rangle = U |A\rangle,
\label{e:ba}
\end{equation}
it follows that $|B\rangle $ and $|A\rangle$ are degenerate, since
\begin{equation}
E_A = \langle A|H|A\rangle = \langle B|H|B\rangle = E_B.
\label{e:eqs}
\end{equation}
In order that this degeneracy manifest itself, however, it is necessary that
the ground state of the system be invariant under such a transformation.
Writing $|A\rangle$ and $|B\rangle$ in terms of creation operators,
\begin{equation}
|A\rangle = \phi_A|0\rangle \quad\quad \rm{and}\quad\quad |B\rangle=\phi_B|0\rangle
\label{e:eee}
\end{equation}
with 
\begin{equation}
U\phi_A U^\dagger = \phi_B,
\label{e:phirel}
\end{equation}
one sees that Eq.(\ref{e:ba}) holds only if 
\begin{equation}
|0\rangle = U|0\rangle,
\label{e:gs}
\end{equation}
i.e. the ground state is invariant under the symmetry group.   Should this
{\it not} be the case, one speaks of a {\it spontaneously broken symmetry}.

Denoting $U$ as $U=\exp(i\varepsilon^aQ^a)$ in terms of the (continuous) group
parameters $\varepsilon^a$ and the generators of the symmetry
\begin{equation}
Q^a = \int d^3x J^a_0(x),
\label{e:qs}
\end{equation}
the statement Eq.(\ref{e:gs}) is seen to coincide with the equivalent form
\begin{equation}
Q^a|0\rangle \ne 0
\label{e:alt}
\end{equation}
although 
\begin{equation}
[Q^a,H] = 0.
\label{e:comm}
\end{equation}
The direct consequence of this statement is that $HQ^a|0\rangle=0$, or that
there must exist a spectrum of massless particles with quantum numbers
specified by the generators of the symmetry.    This constitutes the 
{\it Goldstone theorem}.     To be more precise, one can formulate this
as follows:
 given that a Hamiltonian
has continuous symmetries described by groups $G_1$ requiring $N_{G_1}$
generators, while the ground state is invariant under groups $G_2$ requiring
$N_{G_2}<N_{G_1}$ generators, the spontaneous breakdown of chiral symmetry
leads to the existence of $N_{G_1}-N_{G_2}$ Goldstone bosons
\cite{gold}.

Let us investigate now how this is applied to QCD.

\subsection{Chiral symmetry in QCD}

In this section, we analyse the symmetries of quantum chromodynamics, and 
compare this with the symmetry of the vacuum, determined purely by viewing
the experimental spectrum.    Start by examining the QCD
 Lagrangian itself, which  
 can be written in a compact fashion as
\begin{equation}
{\cal L}_{QCD} = \bar\psi(i\not\!\! D - m_0)\psi - \frac 14 tr_c
G_{\mu\nu}^a
G^{\mu\nu}_a,
\label{e:lqcd}
\end{equation}
where $G_{\mu\nu}$ is the field strength tensor of the gluon field,
\begin{equation}
G_{\mu\nu}^a = \partial_\mu G_\nu^a - \partial_\nu G_\mu^a -gf_{abc}
G_\mu^b G_\nu^c,
\label{e:fieldstr}
\end{equation}
$D^\mu$ is the covariant derivative,
\begin{equation}
D_\mu = \partial_\mu  + ig(\frac 12\lambda_a) G_\mu^a(x) 
\label{e:covder}
\end{equation}
and  $f_{abc}$ are the structure constants
of the symmetry group SU(3) \cite{itzyk}.
 The quark field is a \emph{vector} in flavor
space,
\begin{equation}
\psi = \left(\begin{array}{c} \psi_u(x)\\ \psi_d(x)\\ \psi_s(x)\\ . \\ . \\ .
\\
\end{array} \right)
\label{e:psis}
\end{equation}
and the (current) quark mass matrix is a diagonal matrix in flavor space,
\begin{equation}
m_0 = diag[m_0^u, m_0^d, m_0^s,...],
\label{e:massmatr}
\end{equation}
so that the second term in Eq.(\ref{e:lqcd}) is
\begin{equation}
\bar\psi m_0\psi = \sum_f m_0^f\bar\psi_f\psi_f.
\label{e:break}
\end{equation}
If the quarks are massless, then the Lagrangian Eq.(\ref{e:lqcd}) contains
no term
of the form Eq.(\ref{e:break})
which  can mix left and right handed components of the quark fields,
that are defined as
\begin{equation}
\psi_{R,L} =\frac 12(1\pm\gamma^5)\psi,
\label{e:leftright}
\end{equation}
i.e. these two fields are independent, and the Lagrangian remains invariant
under transformations that individually transform these fields,
\begin{equation}
\psi_{R,L} \rightarrow U_{R,L}\psi, \quad\quad U_{R,L} \in U(N_f),
\label{e:transf}
\end{equation}
and these are called chiral symmetries.   However,    a mass term
of the form Eq.(\ref{e:break}) spoils this
invariance since
\begin{equation}
{\rm Terms} \sim \bar\psi \psi = \bar \psi_L\psi_R + \bar \psi_R \psi_L
\label{e:break2}
\end{equation}
mix left and right handed fields.   Thus the term $m_0\bar\psi\psi$
constitutes an {\it explicit} symmetry breaking.

The QCD Lagrangian Eq.(\ref{e:lqcd}) is invariant under several
transformations, such as
\begin{eqnarray}
\psi\rightarrow\psi' &=& e^{i\alpha}\psi \nonumber \\
\psi\rightarrow\psi' &=& e^{i\alpha\lambda^a}\psi 
\label{e:inv}
\end{eqnarray}
etc.    Accordingly, there are conserved Noether
currents that correspond to these symmetries.   They are
\begin{eqnarray}
V_0^\mu &=& \bar\psi \gamma^\mu\psi \quad\quad A_0^\mu =
\bar\psi\gamma^\mu\gamma^5 \psi \label{e:currs} \\
V_a^\mu &=& \bar\psi\gamma^\mu\frac {\lambda_a}2\psi \quad\quad 
A_a^\mu = \bar\psi\gamma^\mu \gamma_5\frac{\lambda_a}2\psi 
\end{eqnarray}
Among these is the current $A_0^\mu$, which  corresponds to the transformation 
$\psi\rightarrow \psi'=\exp(i\gamma_5\alpha)\psi$,
 where $\alpha$ is a continuous
parameter. However, despite its appearance, this current  is \emph{not}
 conserved,
\begin{equation}
\partial^\mu A_\mu^0 = \frac{N_c}{8\pi^2} tr_c G_{\mu\nu}\tilde G^{\mu\nu}.
\end{equation}
This means that it
 does not reflect an underlying symmetry of the Lagrangian and its
breaking was resolved by 't Hooft as being due to the presence of
instantons \cite{hooft}.

One may thus identify the (continuous) symmetry groups of QCD as being 
generated by the charges of the remaining symmetries, and this is
\begin{equation}
G_1 = SU_L(N_f)\otimes SU_R(N_f) \otimes U_V(1).
\label{e:symmlag}
\end{equation}
On the other hand, by examining the particle spectrum that is observed
experimentally, one finds that the symmetry of the vacuum is 
\begin{equation}
G_2 = SU_V(N_f)\otimes U_V(1).
\label{e:symvac}
\end{equation}
Accordingly, there must be $N_f^2-1$ massless Goldstone particles and these
have the quantum numbers obtained from applying the axial charge operators
to the vacuum, i.e. $J^P=O^-$.    In the case of two flavors, there are
three such states, which are identified as corresponding to the  charged and 
neutral pions.   For three flavors,
one identifies the eight pseudoscalars as the pions, kaons and eta.  One sees
that the explicit symmetry breaking in this case is larger:   $m_0^s\simeq
150 $MeV in comparison with  $m_0^u\simeq m_0^d\simeq 5 $ MeV.

The phase in which a system finds itself is usually characterized by an 
\emph{order parameter}.    This is an
operator that transforms in a non-trivial fashion under the broken symmetry. 
  Generally order parameters  have the property
of being zero in the symmetric or restored phase and non-zero in the
spontaneously broken phase, but this is not necessarily so.
   There are many possible ways of  choosing an
order parameter. The major criterion for doing so is that 
the order parameter should display that same invariances as the ground
state.    In the case of quantum chromodynamics, the ground
state of QCD is invariant under Lorentz transformations and spatial 
reflections.   The order parameter must thus 
be invariant under these same symmetries,
and as such must be a scalar.   The operator $\bar\psi\psi$ is the simplest
choice. One thus makes the  choice of  $\langle \bar\psi\psi\rangle$, which
is referred to as the quark condensate.

\subsection{Lattice gauge simulations}

Simulations of QCD on the lattice provide the most exact knowledge that we
have of this theory that is derived from the QCD Lagrangian itself.   The
Lagrangian is discretized in space and time dimensions, and the variation
with respect to the temperature of physical quantities is formally 
controlled by varying the size of the lattice in the temporal direction
\cite{karsch,laerman},
since
\begin{equation}
T=1/N_\tau a,
\end{equation}
where $a$ is the lattice size and $N_\tau$ the temporal extent.   In what
follows, we simply list the major results that have been extracted via
this methodology over the past few years. 
   We show the temperature dependence of the chiral
and deconfinement order parameters, discuss critical exponents, meson
screening masses and $U_A(1)$ resotration.   Finally, we show plots of the
bulk thermodynamic quantities.

\subsubsection{Order parameters}  

The following recent results \cite{karsch,laerman}
have emerged from the lattice gauge studies:

\begin{itemize}
\item  The pure gauge sector of QCD displays a well-established first order
chiral transition at a rather high critical value of the temperature,
 $T_c=270(5)$ MeV.   The bulk properties 
for such a system are also well known
\cite{ukawa}.
\item Full QCD including fermions displays a
chiral phase transition at  far lower critical temperature
than that observed for pure gluonic systems.   One finds
$T_c\simeq 150$ MeV for two flavors of quark.

\item Studies of the Polyakov loop for quenched QCD places the  critical
temperature determined from the  
order parameter for deconfinement, $T_D$ at about the same temperature at
which  the
chiral transition $T_c$ occurs \cite{hwa}, i.e.
 \begin{equation}
T_D \simeq T_c
\label{e:tsequal}
\end{equation}

\end{itemize}
This can be directly seen from Fig.~1, in which the order parameter for the
chiral and deconfinement transitions are shown, together with their 
susceptibilities, as a function of $\beta=6/g^2$, over the transition
region.   Large (small) values of $\beta$ represent the high (low)
temperature regime.

Based on these points, our physical (but heuristic!)
 understanding of the situation
 is that, at low energies, one has only hadronic states.   These can be 
thought of as  consisting of
 quarks carrying a dynamically generated quark mass $m=m_u=m_d$ for 
two flavors, and constructed into baryonic states or mesonic states 
according to the Goldstone theorem.
At the temperature at which 
 where chiral symmetry is restored, $T_c$, and the
constituent quarks take on their current mass value, deconfinement
occurs simultaneously.    The hadronic states dissolve, and one moves to 
a plasma containing only quark and gluonic degrees of freedom.  

%
%
\begin{center}
\begin{figure}
\epsfig{file=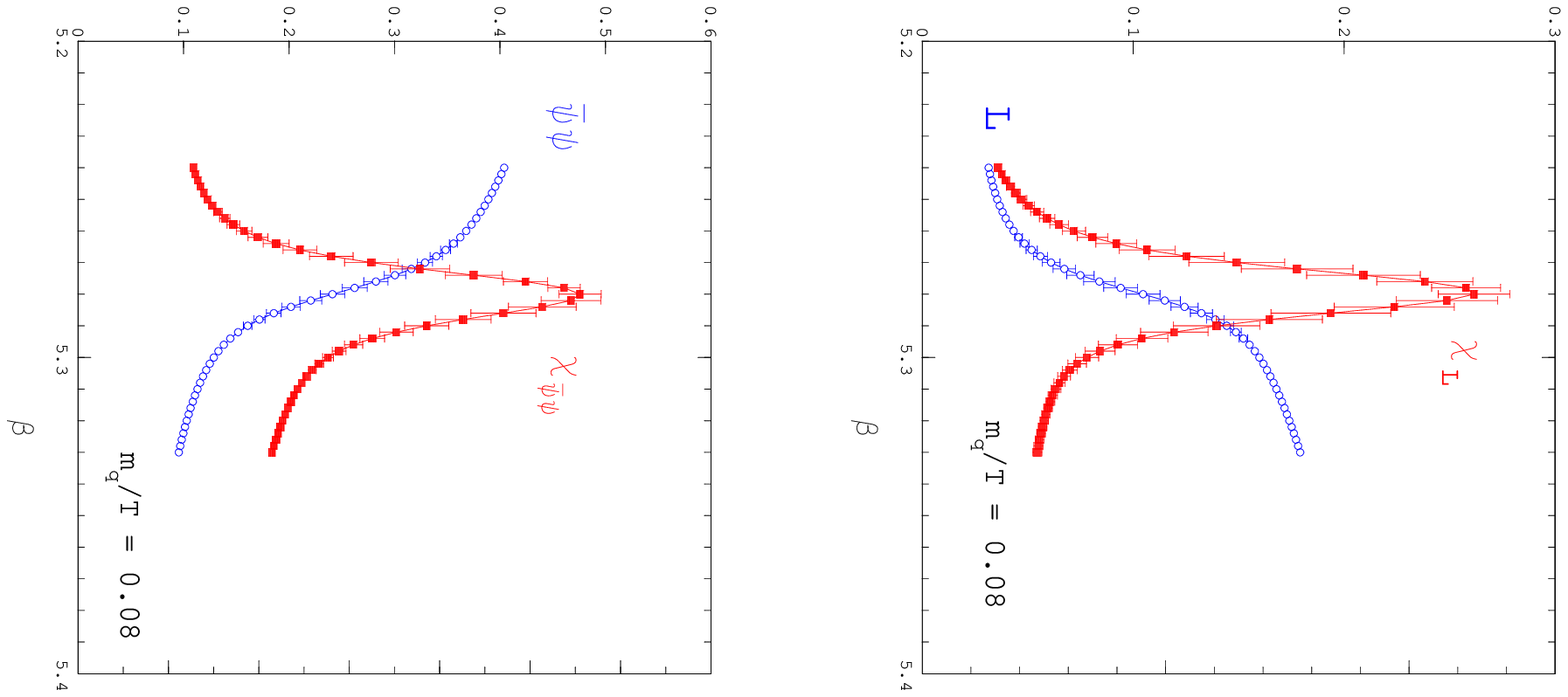,angle=90,scale=0.8}
\caption{Order parameters for the chiral and deconfinement transitions
(lower and upper figures, respectively) are plotted as a function of the
inverse QCD coupling $\beta = 6/g^2$.   High (low) values of $\beta$
correspond to high (low) values of the temperature.
     The associated susceptibilities
are also plotted in each case. Courtesy of \cite{laer}. }
\end{figure}
\end{center}

\subsubsection{Critical Exponents}

An obvious question that one may pose, when faced with a phase transition,
is what are the critical exponents that govern the transition?   Pisarski
and Wilczek suggested that the dynamics of QCD is controlled by an effective
 scalar Lagrangian, constructed along the lines of the linear $\sigma$
model
 \cite{piwi},
and, which for two flavors of quarks,  
has $SU(2)\otimes SU(2)=O(4)$ symmetry. Now, according to arguments 
of universality \cite{golden},
 only the symmetry structure and dimensionality determine
the values of the critical exponents, i.e. one expects that one should
obtain the critical exponents of a 3D $O(4)$ symmetric spin model.
   The task of studying the critical exponents has been
undertaken by a lattice group \cite{laerman}.   
Noting that the masses, which are responsible for {\it explicit} chiral symmetry
breaking, 
 play an analogous role to that of a magnetic field in the
superconducting transition,
 these authors \cite{karsch,laerman} have  adopted the convention of defining
a scaled quark mass as $h=m_q/T$ and the reduced temperature $t=(T-T_c)/T_c$.
With this convention, the free energy density scales as
\begin{equation}
f(t,h) = -\frac TV\ln Z = b^{-1} f(b^{y_t} t, b^{y_h}h),
\label{e:free}
\end{equation}
introducing the thermal $(y_t)$ and magnetic $(y_h)$ critical exponents.
Here $b$ is an arbitrary scaling factor.  There are various scaling
relations that can be derived using Eq.(\ref{e:free}).    In particular,
one can show that the chiral order parameter scales as 
\begin{equation}
\langle \bar\psi\psi\rangle(t,h) = h^{1/\delta}F(z),
\label{e:magscale}
\end{equation}
with $z=th^{1/\beta\delta}$, and the chiral susceptibility,
defined as $\chi_m(t,h) =
\partial \langle \bar\psi\psi\rangle/\partial m$, via
\begin{equation}
\chi_m(t,h) = \frac1\delta h^{\frac 1\delta-1}[F(z) - \frac z\beta F'(z)].
\label{e:chim}
\end{equation} 
The more familiar
 critical exponents $\delta$ and $\beta$ are related to $y_t$ and $y_h$
via
\begin{equation}
\beta = \frac{(1-y_h)}{y_t}\quad\quad {\rm and}\quad\quad
 \delta = \frac{y_h}{1-y_h}.
\label{e:betadelta}
\end{equation}
The heights of the peaks of the susceptibilities scale themselves with the
behavior
\begin{equation}
\chi_m^{{\rm peak}} \sim m^{-z_m}\quad\quad{\rm and}\quad\quad 
 \chi_t^{{\rm peak}} \sim
m^{-z_m}
\label{e:peaks}
\end{equation}
with $z_m=2-1/y_h$ and $z_t=(y_t-1)/y_h +1$.

\begin{table}[ht]
\begin{center}
\begin{tabular}{|l c c | c c|} \hline
 & $y_h$ & $y_t$ & $z_m$ & $z_t$ \\ \hline\hline
$O(4)$ & 0.83 & 0.45 & 0.79 & 0.34 \\
$O(2)$ & 0.83 & 0.50 & 0.79 & 0.39 \\
MF & $3/4$ & $1/2$ & $2/3$ & $1/3$ \\ \hline
\end{tabular}
\end{center}
\caption{Critical exponents for $O(4)$, $O(2)$ and mean field theory (MF).
Taken from \cite{laerman}.}
\end{table}

The expected values for the critical exponents for the case of $O(4)$ symmetry,
$O(2)$ symmetry, and mean field exponents (MF) are listed in Table I,
in the form of $y_h$, $y_t$, and the corresponding values of $z_m$ and
$z_t$.
The $O(2)$ symmetry exponents are also listed, because at finite lattice
spacing, the exact chiral symmetry of the staggered fermion action is
$U(1)\simeq O(2)$.    Only sufficiently close to the continuum limit does
one expect to find $O(4)$ exponents.

The calculated results for 
the exponents themselves, evaluated on different spatially
sized lattices, are summarized in Table~2.     Comparing Tables~1 and 2, one
sees that at this stage, no definitive statement about the symmetry of the
underlying Lagrangian can be made from lattice gauge theory.    This is an
indicator that vital study in this field is still necessary to
determine the underlying symmetry group conclusively.
  It is probably necessary to increase
the lattice sizes and move to smaller masses.

\begin{table}[ht]
\begin{center}
\begin{tabular}{|l| c c c|} \hline
 & $8^3$ & $12^3$ & $16^3$ \\ \hline\hline
$z_m$ & 0.84(5) & 1.06(7)& 0.93(8) \\
$z_t$ & 0.63(7) & 0.94(12) & 0.85(12) \\\hline 
\end{tabular}
\end{center}
\caption{Critical exponents, as a function of the lattice size.   Taken from 
Ref.\cite{laerman}.}
\end{table}

\subsubsection{Meson screening masses and $U_A(1)$ restoration}

One of the questions that has raised some theoretical interest in the
last few years is whether the $U_A(1)$ symmetry, i.e. the symmetry 
$\psi\rightarrow e^{i\alpha\gamma_5}\psi$, which leads to the 
{\it non-conserved current} $A_0^\mu$ that is given in Eq.(\ref{e:currs})
is also restored at finite temperature, at some point.    For three flavors,
this occurs trivially.   A demonstration of this, following Ref.\cite{birse}
 is given.

In SU(3), the statement that $U_A(1)$ is restored, implies that $m_\pi=
m_{\eta'}$.    Since the masses of the particles are determined from the
vacuum expectation values of the appropriate meson-meson correlators, we
need to show only that
\begin{equation}
\langle\phi_3(x)\phi_3(0)\rangle = \langle\phi_0(x)\phi_0(0)\rangle,
\label{e:phi3300}
\end{equation}
where $\phi_3(x) = \bar\psi(x) i\gamma_5\lambda_3\psi(x)$ is the correlator
for the $\pi_0$ and $\phi_0(x) =\bar\psi i\gamma_5\lambda_0\psi(x)$ is 
that for the $\eta'$.   If one considers the specific axial transformation
\begin{equation}
\psi(x) \rightarrow\psi'(x) = e^{i\gamma_5(\sqrt 3\lambda_8 -\lambda_3)\frac
\pi 4}\psi(x),
\label{e:trans}
\end{equation}
then, after a little algebra, one finds that the composite fields transform
as
\begin{equation}
\phi_3(x) \rightarrow\phi'{}_3(x) = \sqrt{\frac23}\phi_0(x) +
\sqrt{\frac13}\phi_8.
\label{e:phitrans}
\end{equation}
The correlator composed  of these composite fields itself then transforms as
\begin{eqnarray}
\langle\phi_3(x) \phi_3(0)\rangle&\rightarrow& \langle
\phi'{}_3(x)\phi'{}_3(0)\rangle \nonumber \\
&=& \frac 23\langle\phi_0(x)\phi_0(0)\rangle + \frac 13
\langle\phi_8(x)\phi_8(0)\rangle\nonumber \\
&+&  \frac{\sqrt 2}3\langle\phi_0(x)\phi_8(0)\rangle
+ \frac{\sqrt 2}3\langle\phi_8(x)\phi_0(0)\rangle\nonumber \\
\label{e:mult}
\end{eqnarray}
The last two terms of this expression vanish, since the system is 
assumed to be $SU_V(3)$ symmetric.   In addition, this implies that
$\langle \phi_3(x)\phi_3(0)\rangle = \langle\phi_8(x)\phi_8(0)\rangle$,
so that Eq.(\ref{e:mult}) implies that
\begin{equation}
\langle\phi_0(x)\phi_0(0)\rangle = \langle\phi_3(x)\phi_3(0)\rangle,
\label{e:answer}
\end{equation}
or that $m_{\eta'}=m_\pi$.

In retrospect, it is simple to understand why the symmetry must be restored.
Noting that mathematical constructions containing
 traces of fields {\it preserve} the symmetry, while determinants
or antisymmetric functions violate it, one sees that the lowest order
combination of fields that would violate $U_A(1)$ would involve the
completely antisymmetric tensor, and consequently contain {\it three}
field combinations.   Since one requires here only {\it two} field
operator combinations in order to construct a meson-meson correlator, this
must be $U_A(1)$ invariant in the chirally restored phase.   This leads to
the definitive statement:   for $n<N_f$, all $n$-point functions in the
chirally restored phase are $U_A(1)$ invariant.

From the previous argument, it is evident that in SU(2) the situation is
more complicated.   There are two independent chiral multiplets in this case:
$(\sigma,\vec \pi)$ and $(\eta',\vec a_0)$.   In lattice studies, the 
behavior of the masses of the $\pi$ and the $a_0$ have been calculated.
Here the integral over the correlators has been studied,
\begin{equation}
\chi_{M_i} = \int d^4x \langle\phi_i(x)\phi_i(0)\rangle
\label{e:intcorr}
\end{equation}
for $M_i=\pi$ or $a_0$, and the leading behavior of these correlators is
assumed to be $\chi_{M_i} \sim m_{M_i}^{-2}$.   A plot of the
``screening masses'' obtained in this fashion
 is shown in Fig.~2 as a function of $6/g^2$, 
with $g$ the coupling in the QCD Lagrangian, which again represents
increasing temperature over the region of the phase transition.
It is interesting to note that the $\pi$ and $\sigma$ have become 
degenerate:   in this picture, this occurs at some temperature slightly
larger than $T_c$, with the $\sigma$ undershooting the $\pi$ curve and
approaching it from below.     That the $\sigma$ meson undershoots the
$\pi$ curve is not expected from model calculations and may be a lattice
artifact.   This will be discussed in the following sections.
  One sees in Fig.~2 that the mass of the other scalar, the $a_0$,
drops with temperature, but not as drastically as does the $\sigma$.   One
observes 
that  it does \emph{not}\ become degenerate with the $\pi$
and $\sigma$ over the temperature range indicated.    Thus it does not
appear 
from this particular calculation 
that $U_A(1)$ symmetry is restored in SU(2).   An alternate approach, 
however, using the scaling arguments of Brown and Rho \cite{brown}, {\it does}
\ however indicate a degeneracy at the transition temperature \cite{laerman}.
   Thus, in this
section once again, the question of the restoration of $U_A(1)$ symmetry
is not resolved.
 

\begin{center}
\begin{figure}[t]
\epsfig{file=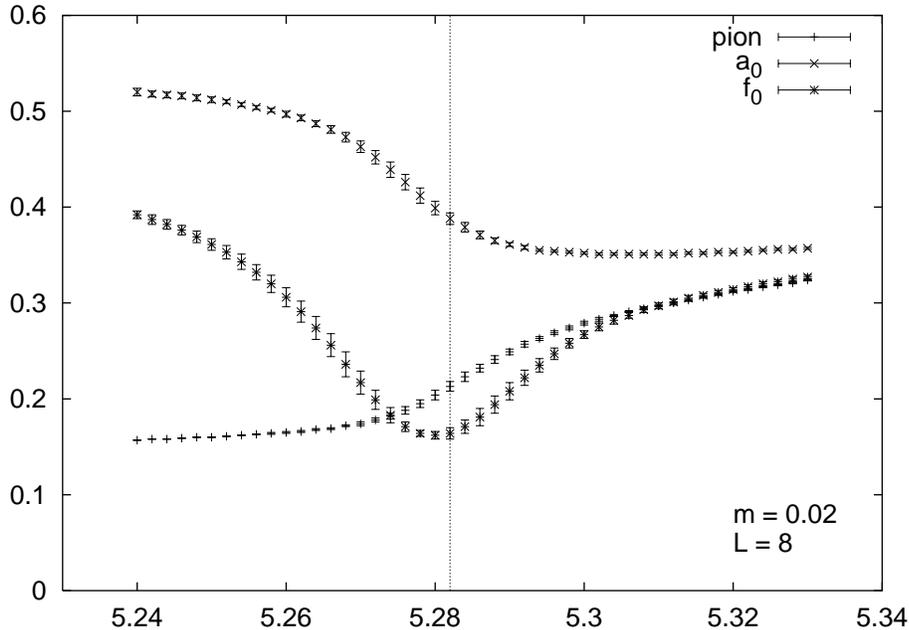}
\caption{Meson screening masses plotted as a function of $\beta=6/g^2$.
The critical value $\beta=\beta_c$ is indicated by the vertical line.}
\end{figure}     
\end{center}

\subsubsection{Bulk thermodynamic quantities.}

One of the most important contributions that lattice physics is able to 
provide are calculations of bulk thermodynamic quantities.   In particular,
the energy density and pressure densities are given by
\begin{equation}
e=\frac {T^2}V\frac \partial{\partial T} \ln Z 
\end{equation}
and
\begin{equation}
p = T\frac{\partial}{\partial V} \ln Z 
\end{equation} 
in terms of the partition function $Z$.
In practice \cite{engels}, the pressure density is obtained from integration
of the difference of the action densities at zero and finite temperature,
\begin{equation}
\frac p{T^4}|_{\beta_0}^{\beta} = N_\tau^4\int_{\beta_0}^\beta d\beta 
(S_0-S_T).
\end{equation}
Note that this quantity is defined in such a way that $p/T^4_{T=0}=0$,
in contrast to setting the usual thermodynamic limit of Nernst, i.e.
the entropy ${\cal S}(T=0)=0$ \cite{yuki}.    While this does not affect
anything that follows, one should bear this in mind when making model 
comparisons, as will be done in Section 3.

In the following figures, we have chosen to illustrate the pressure and
energy densities for lattice simulations that include quark degrees of 
freedom, rather than simply quenched QCD.    In Fig.~3,  we show the
pressure density, plotted as a function of the scaled temperature, for
four flavor QCD on a $16^3\times 4$ lattice.   A comparison is made
on varying the quark masses, and using quenched QCD, in the latter case
with appropriate scaling of the number of degrees of freedom.   One sees
that there is a {\it sharp} rise in the pressure density at $T=T_c$,
and the curve tends to the Stefan-Boltzmann limit, but does not reach it
over the temperature range $(3.5T_c)$ shown.    The deviation from the ideal
gas limit appears to be too large to be described by perturbation theory,
suggesting here that the perturbative regime occurs for
temperatures $T>>T_c$. 


\begin{center}
\begin{figure}[t]
\epsfig{file=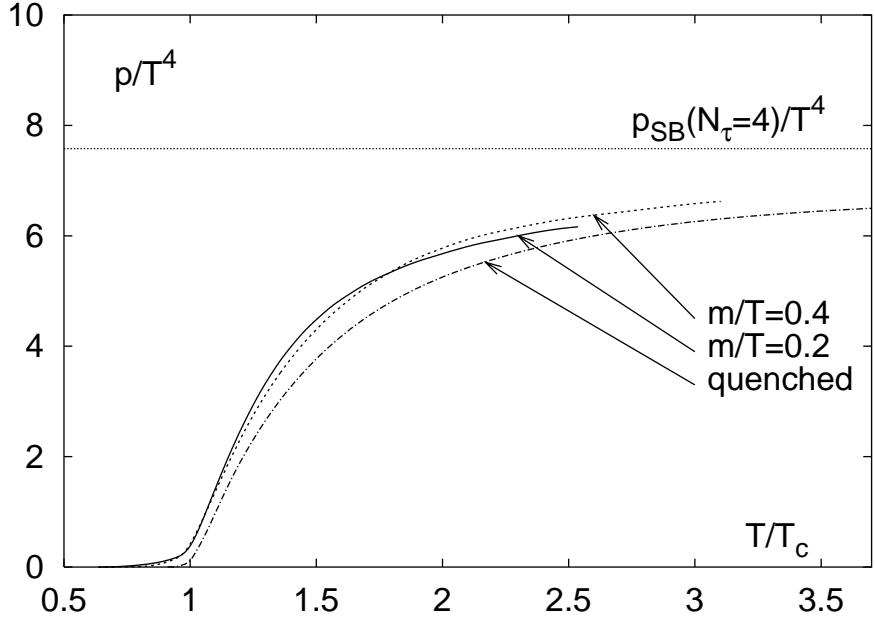}
\caption{Pressure density, plotted as a function of the scaled temperature
$T/T_c$. (Taken from \cite{karsch}.)}
\end{figure}  
\end{center}

The energy density of four flavor QCD on a $16^3\times 4$ lattice is shown
in Fig.~4.  In this case, the energy density remains close to the ideal gas
limit at temperatures of the order of $3T_c$, but overshoots it and
approaches it from above for finite values of the quark mass.    Whether 
this is a lattice artifact or not is presently unclear.


\begin{center}
\begin{figure}[t]
\epsfig{file=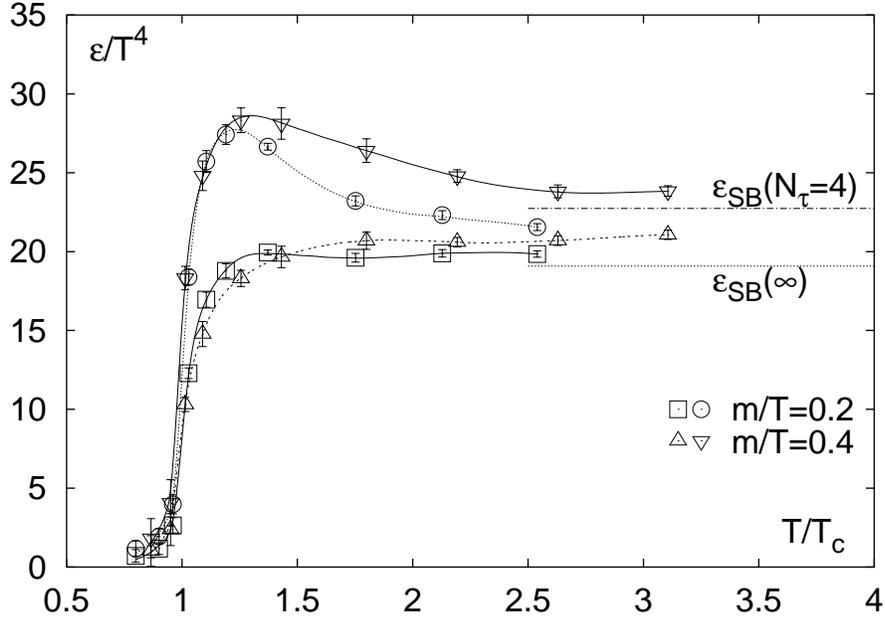}
\caption{Energy density
plotted as a function of the scaled temperature, $T/T_c$. (Taken from
\cite{karsch}.)}
\end{figure}  
\end{center}

In concluding this section, we see that lattice gauge simulations are
reaching a point where one may obtain ``exact'' results that stem directly 
from the discretized QCD Lagrangian.   These can be used as a guide
for constructing simple models, and conversely, simple models and simple
predictions based solely on symmetry considerations such as discussed
here\footnote{Chiral random matrix theory \cite{book} also falls into this
category.}
 may be used as a guide
for the interpretation of the numerical results.   
As has emerged here, there are still many questions that are open for study.

   With this, we turn to a different approach which is regarded
by its protagonists as being an exact low energy representation of QCD,
viz. chiral perturbation theory and investige what is known at finite 
temperature.

\subsection{Chiral perturbation theory}

\subsubsection{A brief introduction}

Chiral perturbation theory starts with the premise that an effective 
Lagrangian for QCD at low temperatures can be written solely in terms of the
observed baryonic (here mesonic) degrees of freedom, in such a way that
global chiral symmetry is enforced.  This is done in its most general form
by collecting the mesonic degrees of freedom into the field
\begin{equation}
U(x) = \exp(i\pi^a\tau_a/F),
\label{e:u}
\end{equation}
where $\pi^a$ are the $SU(2)$ pion fields, $\tau_a$ the 
Pauli matrices, and $F$ the pion decay constant, and contructing a Lagrangian
density that is ordered in momenta.   Such an expansion for the Lagrangian
only starts at $O(p^2)$, and must contain an even number of derivatives in 
order to be Lorentz invariant.   Writing
\begin{equation}
{\cal L}_{QCD}\rightarrow {\cal L}_{eff}
 = {\cal L}_{eff}^{(2)} + {\cal L}_
{eff}^{(4)} + {\cal L}_{eff}^{(6)} + \dots,
\label{e:chptlag}
\end{equation}
the lowest leading order term is
\begin{equation}
{\cal L}_{eff}^{(2)} = \frac 14 F^2 {\rm tr} (\partial_\mu U\partial^\mu
U^\dagger),
\label{e:leff2}
\end{equation}
which, taken on its own, is the (non-renormalizable) sigma model.   QCD,
as we have already discussed,  is
however {\it not} completely invariant under chiral symmetry.  There is
an explicit breaking of the symmetry due to the presence of the current
quark mass matrix.    The symmetry breaking term is in general given as
\begin{equation}
{\cal L}_{sb} = f(U,\partial U, \dots) \times m^0,
\label{e:lsb}
\end{equation}
where $m^0$ is the (real and diagonal) current quark mass matrix.
One incorporates this into the effective Lagrangian by making not only
an expansion in powers of the derivatives, but also in powers of $m^0$,
i.e. ${\cal L}_{sb}\sim f(U)\times m^0$ to leading order.  More precisely,
this term takes the form (that is Lorentz and parity invariant)
\begin{equation}
{\cal L}_{sb} = \frac 12 F^2 B {\rm tr}(m^0(U+U^\dagger)),
\label{e:lsbprec}
\end{equation}
introducing the new constant $B$.  This is generally included in the
definition of ${\cal L}_{eff}^{(2)}$, i.e.
\begin{equation}
{\cal L}_{eff}^{(2)} = \frac14 F^2{\rm tr}(\partial_\mu U\partial^\mu U^\dagger)
+ \frac 12 F^2 B {\rm tr}(m^0(U+U^\dagger)).
\label{e:l2full}
\end{equation}
In this reckoning, one can thus regard $m^0$ as being of $O(p^2)$.

To make physical sense of the constant $B$, one may expand the field
$U=\exp(i\pi\cdot\tau/F)$ in powers of the pion field $\pi$.    The 
symmetry breaking part of the Lagrangian then becomes
\begin{equation}
{\cal L}_{sb} = (m_u^0 + m_d^0) B[F^2 - \frac 12 \pi^2 +\frac 1{24} \pi^4
F^{-2} + 
\dots ]
\label{e:expanded}
\end{equation}
The first term in this expansion gives the vacuum energy generated by the
symmetry breaking.   The second term generates the pion mass, while the 
higher order terms describe further interactions of the $\pi$ fields.   By
direct analogy with the QCD Hamiltonian, we know that the derivative of
$H_{QCD}$ with respect to $m_u^0$ generates the operator $\bar u u$.   Thus
the derivative of the vacuum energy with respect to the current quark mass
gives the vacuum expectation value of this operator.   Applying this to 
${\cal L}_{eff}$, one has
\begin{equation}
\langle 0|\bar u u|0\rangle = \langle 0 |\bar d d|0\rangle = -F^2B\{1+
O(m)\} ,
\label{e:condu}
\end{equation}
indicating that $B$ is related to the condensate.   Since the pion mass is 
given as 
\begin{equation}
m_\pi^2 = (m_u^0 + m_u^0)B\{1+O(m)\},
\label{e:mpiexpand}
\end{equation}
one obtains the Gell-Mann-Oakes-Renner (GOR) relation \cite{gor},
\begin{equation}
F_\pi^2M_\pi^2 = (m_u^0+m_d^0)|\langle0|\bar u u|0\rangle|
\label{e:gmor}
\end{equation}
from Eqs.(\ref{e:condu}) and (\ref{e:mpiexpand}), on eliminating $B$.

To order $p^4$, the effective Lagrangian would contain two additional
independent terms
in the event that no current quark mass were present, i.e.
one would include two new terms
\begin{equation}
{\cal L}_{eff}^{(4)} = \frac 14l_1({\rm tr}\{\partial_\mu U^\dagger\partial
^\mu U\})^2 + \frac 14 l_2{\rm tr} (\partial_\mu U^\dagger\partial_\nu U)
{\rm tr}(\partial^\mu U^\dagger\partial^\nu U).
\label{e:l4eff}
\end{equation}
with new low energy constants $l_1$ and $l_2$.
Including the current quark mass matrix again to construct an explicit
symmetry breaking terms 
requires the inclusion of
further additional
terms, as was the case for ${\cal L}^{(2)}_{eff}$.
For most purposes, this is sufficient.   However, to obtain the most general
form from which all propagators can be derived, it is useful to introduce
\emph{external fields} into the Lagrange density.
 Here the essential additions are $v_\mu(x)$ and $a_\mu(x)$
that are vector and axial vector in nature and which can be regarded as
being of order $p$.     Then, using the original notation of Ref.\cite{chpt},
 the complete
set of terms that contribute to ${\cal L}_{eff}^{(4)}$ were worked out by
these authors and found to be, for SU(3)
\begin{eqnarray}
{\cal L}_{eff}^{(4)} &=& L_1\langle\nabla_\mu U^\dagger\nabla^\mu U\rangle^2
+ L_2\langle\nabla_\mu U^\dagger\nabla_\mu U\rangle\langle\nabla^\mu
U^\dagger\nabla^\nu U\rangle \nonumber \\
&+&L_4\langle \chi U^\dagger+\chi^\dagger U\rangle\langle\nabla
U^\dagger\nabla^\mu U\rangle + L_6\langle\chi U^\dagger + \chi^\dagger
U\rangle^2 \nonumber \\
&+& L_8\langle\chi U^\dagger \chi U^\dagger+ U\chi^\dagger U\chi^+\rangle
+ L_{10}\langle U^\dagger F^{\mu\nu}_R U F_{L\mu\nu}\rangle\nonumber \\
&+& H_1\langle F^{\mu\nu}_R F_{R\mu\nu}
 + F^{\mu\nu}_L F_{L\mu\nu}  
+
H_2\langle \chi^\dagger\chi\rangle
+ L_7\langle\chi U^\dagger - U\chi^\dagger\rangle^2 \nonumber \\
 &+& L_5\langle\nabla_\mu
U\nabla^\mu U^\dagger(\chi U^\dagger +
U\chi^\dagger)\rangle
+ iL_9\langle F^L_{\mu\nu}\nabla^\mu \nabla^\nu U^\dagger + F^R_{\mu\nu}
\nabla^\mu\nabla^\nu U\rangle
\nonumber \\
&+& L_3 \langle \nabla_\mu U\nabla^\mu U^\dagger \nabla_\nu U \nabla^\nu.
U^\dagger\rangle
\label{e:l4efflong}
\end{eqnarray}
where
using a different notation to Eq.~(\ref{e:l4eff}) now, the low energy
constants $L_1$ to $L_{10}$, and $H_1$ and $H_2$ have been introduced.  The
angular brackets are a shorthand notation for the trace.
In this expression, one notes that the covariant derivative that is
constructed using the external field must now appear,
\begin{equation}
\nabla_\mu = \partial_\mu - i\{a_\mu,U\},
\label{e:covderchpt}
\end{equation}
and $F^{\mu\nu}$ is the field strength tensor constructed from the {\it 
external}
 field, i.e.
\begin{equation}
F^{\mu\nu}_{R,L} = \pm\partial^\mu a^\nu \mp \partial^\nu a^\mu - i[
a^\mu,a^\nu].
\label{e:fmunutensor}
\end{equation}
Terms involving the current quark mass have been summarized into the field
$\chi = 2 B\hat m$, with $\hat m = (m^u_0 + m^d_0)/2$.  
Note that the low energy constants $L_i$ 
 become renormalized when physical
quantities are calculated, as this theory is perturbatively renormalizable
\emph{order by order}.
    A certain number of such physical quantities that are measured in
experiment  must then be
used to fit the renormalized parameters at a given mass scale.
Given definite values for these constants, predictions 
of other quantities can then be made. 

  Three ingredients
are  essential to any application that attempts to calculate quantities
for chiral perturbation theory to a specific order.   For example, should
one
wish to calculate  to $O(p^4)$, the following steps must be taken:
(1) The general ${\cal L}^{(2)}_{eff}$ of order $p^2$ is to be used at
both the tree and one loop level. (2) The general ${\cal L}^{(2)}_{eff} $
of order $p^4$ is to be used only at tree level.   (3) A renormalization 
program must be implemented to make physical predictions.   The extension of
this procedure to higher powers in $p^2$ is obvious.

Let us look at a standard example for the derivation of the pion mass
\cite{donoghue}.   In what follows, we denote the low energy  
constants appropriate to SU(2) \footnote{These can be simply related to the
$l_i$ of Eq.~(\ref{e:l4eff}),
 and the reader is referred to \cite{gasser} for explicit details.}
two flavors as being $L_i^{(2)}$. 
If one
expands the Lagrangians ${\cal L}_{eff}^{(2)}$  and ${\cal L}_{eff}^{(4)}$
in terms of the pion fields, one finds
\begin{equation}
{\cal L}^{(2)}_{eff} = \frac 23 [\partial^\mu\pi\partial_\mu\pi - m^2\pi
\cdot \pi] + \frac {m^2}{6F^2}[(\pi\cdot\partial^\mu\pi)(\pi\cdot
\partial_\mu\pi)
-(\pi\cdot\pi)(\partial^\mu\pi\cdot\partial_\mu\pi)] + O(\pi^6),
\label{e:exl2}
\end{equation}
while
\begin{eqnarray}
{\cal L}^{(4)}_{eff} &=& \frac{m^2}{f^2}[16 L_4^{(2)} + 8L_5^{(2)}]
\frac 12\partial_\mu\pi\cdot\partial^\mu\pi \nonumber \\
&-& \frac{m^2}{F^2}[
32 L_6^{(2)}+16L_8^{(2)}]\frac12\hat m^2\pi\cdot\pi + O(\phi^4).
\label{e:l4expand}
\end{eqnarray}
The terms in ${\cal L}^{(4)}_{eff}$ that are of order $\pi^4$
contribute to physical quantities via one loop diagrams and one therefore 
does not need to consider these in a calculation to order $O(p^4)$.
What is required however, are the one loop diagrams that are generated by
${\cal L}_{eff}^{(2)}$.  For a calculation of the
the renormalized  pion mass, however, one
can avoid 
evaluating any diagrams at all  by simply considering
all possible contractions of two fields in these terms in 
${\cal L}^{(4)}_{eff}$, to arrive at an ``effective'' effective
Lagrangian, that takes the form
\begin{eqnarray}
{\cal L}^{(4)}_{eff} &=& \frac 12\partial^\mu\pi\cdot\partial_\mu\pi
-\frac 12 m^2\pi\cdot\pi +\frac{5m_\pi^2}{12m^2}I(m_\pi^2)\pi\cdot\pi
\nonumber \\  
&+&\frac 1{6F^2}(\delta_{ik}\delta_{jl} - \delta_{il}\delta_{kl})I(m^2_\pi)
(\delta_{ij}\partial^\mu\pi_k\partial_\mu\pi_l +
\delta_{kl}m_\pi^2\pi_i\pi_j) \nonumber \\
&+& \frac 12\partial_\mu\pi\partial^\mu \pi\frac{m_\pi^2}{F_\pi^2} [16
L_4^{(2)} + 8L_5^{(2)}] - \frac 12 m_\pi^2
\pi\cdot\pi\frac{m_\pi^2}{F_\pi^2}
[32 L_6^{(2)}+16L_8^{(2)}].\nonumber \\
\label{e:effeff}
\end{eqnarray}
In obtaining this result, the Feynman propagator 
\begin{equation}
i\Delta_{Fjk}(0) = \langle
0|\pi_j(x)\pi_k(x)|0\rangle=\delta_{jk}I(m_\pi^2)
\label{e:feyn}
\end{equation}
has been introduced and is written in terms of the integral
\begin{equation}
I(m_\pi^2) = \mu^{4-d}\int\frac{d^dk}{(2\pi)^d} \frac i{k^2-m_\pi^2}
=\mu^{4-d}(4\pi)^{d/2}\Gamma(1-\frac d2)(m_\pi^2)^{\frac d2-1}
\label{e:int}
\end{equation}
that is treated with dimensional regularization, $d$ being an arbitrary
dimension.   In addition, use has been made of the fact that derivatives
of the Feynman propagator, defined as 
\begin{equation}
-\partial_\mu\partial_\nu i \Delta_{Fjk}(0) = \langle 0|\partial_\mu
\pi_j(x)\partial_\nu\pi_k(x)|0\rangle = \delta_{ij} I_{\mu\nu}(m_\pi^2)
\label{e:intder}
\end{equation}
can be expressed in terms of the integral $I(m_\pi^2)$ via
\begin{equation}
I_{\mu\nu}(m_\pi^2) = \mu^{4-d}\int\frac{d^dk}{(2\pi)^d}k_\mu k_\nu \frac
i{k^2-m_\pi^2} = g_{\mu\nu} \frac{m_\pi^2}dI(m_\pi^2).
\label{e:imunu}
\end{equation}
Regrouping the kinetic and mass terms, Eq.(\ref{e:effeff}) becomes
\begin{eqnarray}
{\cal L}_{eff} &=& \frac 12\partial^\mu \pi\cdot\partial_\mu \pi [
1+ (16 L_4^{(2)} + 8 L_5^{(2)}) \frac{m_\pi^2}{F_\pi^2} - \frac
{2}{3F_\pi^2} I(m_\pi^2)] \nonumber \\
&-&\frac12 m^2\pi\cdot\pi[1+(32L_6^{(2)}+16L_8^{(2)})\frac{m_\pi^2}{F_\pi^2}
-\frac{1}{6F_\pi^2} I(m_\pi^2)].
\label{e:regroup}
\end{eqnarray}
By expanding this expression in powers of $d-4$ and renormalizing the
pion field as $\pi_r=Z_\pi^{-1/2}\pi$, with
\begin{equation}
Z_\pi = 1-\frac{8m_\pi^2}{F_\pi^2}(2L_4^{(2)} + L_5^{(2)}) +\frac
{m_\pi^2}{24\pi^2F_\pi^2}[\frac2{d-4} + \gamma-1 -\ln 4\pi + \ln\frac
{m_\pi^2}{\mu^2}],
\label{e:z}
\end{equation}
one obtains the canonical form for the effective Lagrangian for  pion fields,
\begin{equation}
{\cal L}_{eff} = \frac 12\partial_\mu\pi_r\partial^u\pi_r - \frac 12
M_\pi^2\pi_r\cdot\pi_r,
\label{e:lcanon}
\end{equation}
with the identification of the physical pion mass as
\begin{equation}
M_\pi^2 = m^2[1+
\frac{m_\pi^2}{32\pi^2F_\pi^2}\ln
\frac{m_\pi^2}{\mu^2} -\frac{8m_\pi^2}{F_\pi^2}L_{comb}].
\label{e:mfantasy}
\end{equation}
Here $L_{comb} = 2L_4^{(2)}{}^r + L_5^{(2)}{}^r - 4L_6^{(2)}{}^r -
2L_8^{(2)}{}^r$.    In the original paper of Gasser and Leutwyler
\cite{chpt},
$M_\pi^2$ was not obtained in this fashion, but rather from the expansion
of the Fourier transform of the axial vector correlator, which has the 
form
\begin{eqnarray}
J_{\mu\nu}^{ik}(p) &=& i\int d^4s e^{ip(x-y)}\langle0|TA_\mu^i(x)
A_\nu^k(y)|0\rangle\nonumber \\
&=& \delta^{ik}\{\frac{p_\mu p_\nu F_\pi^2}{M_\pi^2-p^2} + \dots\},
\label{e:acorr}
\end{eqnarray}
where $A_\mu^i(x) = \bar\psi(x)\gamma_\mu\gamma_5\frac{\tau^i}2\psi(x)$.
From this expression, the corresponding expansion for $F_\pi$ has also been 
obtained.

\subsubsection{Cool chiral perturbation theory}

The evaluation of the condensate density at finite temperature was first
carried out by Gerber and Leutwyler \cite{gerber}.
   In their calculation, which
involves ${\cal L}^{(2)}_{eff}$ and ${\cal L}^{(4)}_{eff}$, they find that
the
first term in the behavior of the condensate with temperature is  
quadratically decreasing, {\it i.e.}
\begin{equation}
\langle\bar q q\rangle = \langle 0|\bar qq|0\rangle_{T=0}[1-\frac
{T^2}{8F^2} - \frac{T^4}{384 F^4} - \frac {T^6}{288 F^6}\ln \frac
{\Lambda_q}{T} + O(T^8)].
\label{e:condt}
\end{equation}
This is a result that has been obtained under the assumption that quarks
are massless, i.e. in the chiral limit.
    $\Lambda_q$ is a scale factor constructed from the
renomalized low energy constants, and is expected to be of the order of
$\Lambda_q= 360..580$ MeV.

In a recent publication, Toublan \cite{toublan}
 has investigated pion static properties
with the aim of obtaining $O(p^6)$ accuracy in all quantities and to then
verify the Gell-Mann--Oakes--Renner (GOR) \cite{gor}
 relation at finite temperature.
To do so, the tree, one loop and two loop diagrams of ${\cal L}^{(2)}_{eff}$
are required, the tree and one loop graphs of ${\cal L}^{(4)}_{eff}$ are
required plus the tree level graphs of ${\cal L}^{(6)}_{eff}$.     In doing
so, the result of Eq.(\ref{e:condt}) has been  reconfirmed.
    In addition, the
mass
$M_\pi(T)$ and pion decay constant as a function of temperature are also
evaluated, using the finite temperature axial vector correlator.    In
total,
thirty-six Feynman graphs contribute to the correlator at this order!  
However,  in
the chiral limit, one is still lucky enough to have 
simple analytic forms  for the temperature
dependence.   One finds
\begin{equation}
\frac{M_\pi^2(T)}{M_\pi} = 1 + \frac {T^2}{24 F^2} - \frac{T^4}{36 F^4}\ln
\frac{\Lambda_M}T + O(T^6),
\label{e:mpit}
\end{equation}
while
\begin{equation}
\frac{Re[F_\pi^t(T)]^2}{F_\pi^2} |_{\hat m=0} = 1-\frac{T^2}{6F^2}
+
\frac{T^4}{36F^4}\ln \frac{\Lambda_T}T + O(T^6).
\label{e:ftpi}
\end{equation}
and
\begin{equation}
\frac{Re[F_\pi^t(T)-F_\pi^s(T)]}{F_\pi}|_{\hat m=0} = \frac {T^4}{27F^4}\ln
{\Lambda_\Delta}T + O(T^6),
\label{e:ftfs}
\end{equation}
where $\Lambda_{M,T,\Delta}$ are various scales, whose sizes are determined
by the renormalized  couplings $L_1^r\dots L_{10}^r$ that are a function of
scale.   They are determined numerically to be
 $\Lambda_M\simeq 1.9$ GeV, $\Lambda_T\simeq
2.3$ GeV, and $\Lambda_\Delta = 1.8$ GeV.    Note that, at finite
temperature, there is a separation of ``temporal'' and ``spatial'' pion decay
constants.  This comes about since Lorentz invariance is not maintained
in a heat bath and the the singular part of the 
 axial two point function takes the
form
\begin{equation}
A_{\mu\nu}(q,T) = -\frac{f_\mu(q,T)f_\nu(q,T)}{q_0^2 - \Omega^2(\vec q,T)}
\label{e:axtemp}
\end{equation}
where
\begin{equation}
f_0(q,T) = q_0F_t(q,T) \quad\quad {\rm }\quad\quad f_i(q,T) = q_iF_s(q,T),
\label{e:f0fi}
\end{equation}
with $i=1..3$, and the decay constants $F_\pi^{s,t}$ are defined as
\begin{equation}
F_\pi^{s,t}(T) = F_{s,t}(\vec q, T)|_{\vec q=0}.
\label{e:fdefn}
\end{equation}
   The GOR relation is modified so as to read \cite{toublan}
\begin{equation}
\lim_{\hat m\rightarrow0}\frac{M_\pi^2(T){\rm Re}[F^t_\pi(T)]^2}
{\hat m\langle \bar qq\rangle_T} = -1 + O(T^6).
\label{e:gor}
\end{equation}
For this reason, we show graphs for $M_\pi^2(T)/M_\pi^2$ and ${\rm Re}[
F_\pi^t(T)]^2/F_\pi^2$, as a function of temperature in Figs.~5 and 6.
The tree level result is given (dotted curve), together with the one
loop computation (upper dashed line in Fig.~5, lower dashed line in Fig.~6)
 and the two loop approximation (solid
curve).   In both of these figures, a non-zero value of the
quark mass has been assumed for these curves.    In the chiral limit, one
finds the lower (upper) dashed curve in Fig.~5 (6).
    What is evident from these two figures, is that
chiral perturbation theory is not converging and 
 appears to provide an {\it oscillating} series for these
quantities.   Thus for larger
values of the temperature, $T>100$ MeV say, one sees that the pion mass
{\it decreases} with temperature in the two loop approximation, in
contradistinction with the one loop result, the lattice 
results of the last section,  and also in contradistinction
with the model results obtained in the Nambu--Jona-Lasinio model, which will
be presented later on in the following section.
    Convergence at temperatures in this
range appears to be problematic, which is perhaps an indication that the
series is at best asymptotic, or changes its nature due to the onset of
the phase transition.  In this range, one expects non-analytic behavior
and it is unreasonable to expect a perturbation analysis to succeed.
These curves clearly indicate that ChPT at finite temperature can at best
be regarded as cool, so that the fundamental behavior at low temperatures
sets a constraint on the finite temperature behavior of would-be effective
models. 

%
%
\begin{center}
\begin{figure}
\epsfig{file=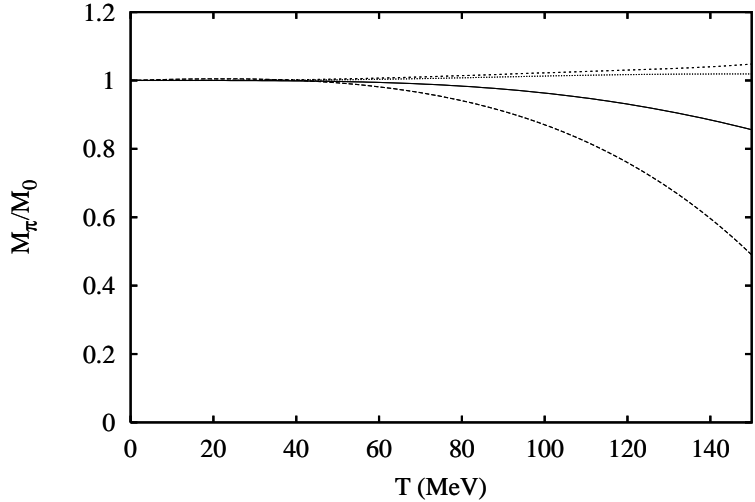,angle=0,scale=0.5}
\caption{The pion mass, scaled by its value at $T=0$ is shown as a function
of the temperature. }
\end{figure}
\end{center}

%
%
\begin{center}
\begin{figure}
\epsfig{file=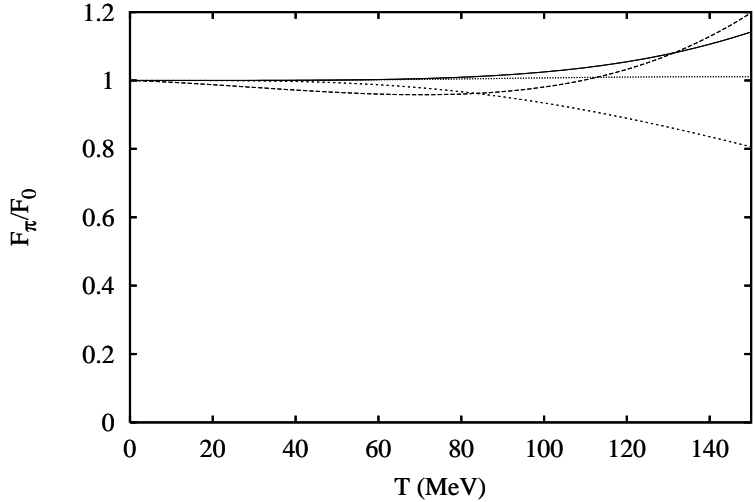,angle=0,scale=0.5}
\caption{The pion decay constant,
 scaled by its value at $T=0$ is shown as a function
of the temperature. }
\end{figure}
\end{center}

\section{The Nambu--Jona-Lasinio model}

The Nambu--Jona-Lasinio (NJL) model has been reviewed in detail by several
authors from different viewpoints \cite{njl,spk,reviews},
 and consequently I do not wish to
present any detail of this model other that a basic introduction here.
Rather the purpose of this chapter is to illustrate that with the simple
equations requiring little computational time, one can reproduce all the
main features of the static properties that have been so arduously extracted
from years of labor on the lattice.   It is extremely encouraging to have
a simple model that can be handled semi-analytically -- one gains a
tremendous amount of insight into the actual functioning of the mechanism
of dynamical symmetry breaking and the consequences thereof.
 
Nevertheless, the NJL model is simply a model -- in contrast to the 
results of the previous section, which are regarded as ``factual'', this
section can only give model-dependent results.   Accordingly it is only
equitable to indicate, in addition to the successes provided by this
approach, the failings also.   These become obvious when examining bulk
thermodynamic properties, such as pressure, energy and entropy densities, 
and will be discussed in what follows.

We shall then turn to dynamic properties, and examine the temperature
dependence of 
scattering amplitudes in the quark-antiquark channel, which displays a
divergence which we term critical scattering, in analogy to the phenomenon
of critical opalescence that is observed in light scattering.

\subsection{Order parameter}

We first consider the order parameter for the chiral transition that is
obtained from the NJL Lagrangian, which, for two flavors of quarks, is
taken to be
\begin{equation}
{\cal L}_{NJL} = \bar \psi(x)(i\not\! \partial - m_0)\psi(x)
       + G [(\bar\psi\psi)^2 + (\bar\psi i\gamma_5\tau \psi)^2],
\label{e:lnjlsu2}
\end{equation}
where $G$ is a dimensionful coupling strength, and $m_0$ denotes the
common current quark mass for $u$ and $d$ quarks.
For three flavors of quarks, we use
\begin{eqnarray}
{\cal L}_{NJL} &=& \bar \psi(x)(i\not\! \partial - m_f^0)\psi(x)
+ G\sum_{a=0}^8 [(\bar\psi \lambda^a\psi)^2 + (\bar
\psi\lambda^a\gamma_5\psi)^2] \nonumber \\
&-& K \{ {\rm det} \bar \psi(1+\gamma_5)\psi + {\rm det} \bar
\psi(1-\gamma_5)\psi \}.
\label{e:su3}
\end{eqnarray}
Here $G$ and $K$ both are 
dimensionful coupling strengths and $m_f^0 = {\rm diag}
(m_0^u, m_0^d, m_0^s)$.   The self-energy, in the mean field approximation,
that corresponds to the lowest order term in an expansion in the inverse
number of colors $N_c$ \cite{quack,lemmer},
 is given as\footnote{Since the coupling strengths
turn out to be large, $G\Lambda^2\sim 2$, an expansion in the number of 
couplings is inadmissable and an alternative expansion scheme must be used.}
\begin{equation}
m = m_0 - 2G\langle\langle \bar\psi\psi\rangle\rangle,
\label{e:msu2}
\end{equation}
and the condensate is given explicitly as
\begin{equation}
\langle\langle \bar\psi\psi\rangle\rangle = \frac {N_cN_f}{\pi^2}
\int_0^\Lambda\frac{p^2}{E_p}[1-f^-(p,\mu) - f^+(p,\mu)],
\label{e:cond}
\end{equation}
with
\begin{equation}
f^\pm(p,\mu) = \frac 1 {[1+\exp\ \beta(E_p\pm\mu)]}.
\end{equation}
One sees that the condensate is directly proportional to the value of the
dynamically generated mass, in the event that the current quark mass is
zero.   Although the situation is more complicated in SU(3), where the
dynamically generated quark masses satisfy coupled equations,
\begin{eqnarray}
m_i&=&m_i^0 - \frac{GN_c}{\pi^2}m_iA(m_i,\mu_i) + \frac{KN_c^2}{\pi^2}
m_jm_kA(m_j,\mu_j) A(m_k,\mu_k), \nonumber \\
&& i\ne j\ne k
\end{eqnarray}
and the function 
\begin{equation}
A(m_f,\mu_f) = \frac {16\pi^2}\beta\sum_n e^{i\omega_n\eta}\int_{|\vec p|<
\Lambda} \frac{d^3p}{(2\pi)^3} \frac 1{(i\omega_n+\mu_f)^2 - E_f^2}
\end{equation}
is proportional to the condensate density for a specific flavor,
\begin{equation}
 A(m_f,\mu_f)\sim \langle\langle\bar\psi\psi\rangle\rangle_f,
\end{equation}
 the dynamically
generated quark masses are equivalently order parameters of the phase 
transition, and we therefore plot these.     They are shown here only
for the SU(3) case, in Fig.~7, for a finite value of the current quark mass
\cite{hadron}.   As expected, the phase transition that occurs in the chiral
limit is washed out and becomes a cross over.
   Another feature that emerges in this model is that
the strange quark mass remains large, even at temperatures $T\sim 300$ MeV, 
and does not reach its current mass value of 150 MeV until $T>>
T_c$.

%
%
\begin{center}
\begin{figure}
\epsfig{file=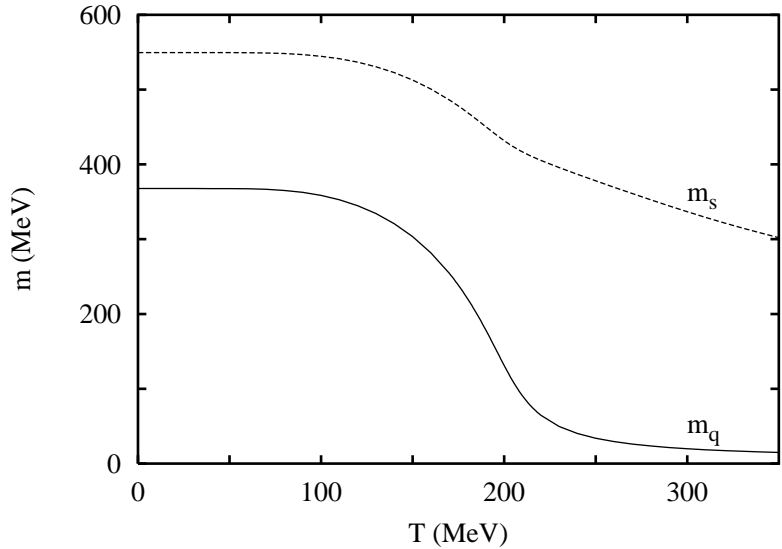,angle=0}
\caption{Temperature dependence of the constituent quark masses.   The solid
line refers to the light quarks, the dashed to the strange quark
\cite{hadron}.}
\end{figure}
\end{center}

\subsection{Meson masses}

The meson masses for the scalar and pseudosalar sectors are determined via
the well-known method of evaluating the quark-antiquark
 scattering amplitude
in the random phase approximation, and searching for poles of this function.
This involves knowing only the irreducible polarization function that one
can construct from a single quark loop, the details of which can be found,
for example,
in \cite{spk,reviews,hadron}.
   One finds the masses that are shown in Figs.~8 and 9
for
the pseudoscalar and scalar sectors, respectively.
In Fig.~8, $2m_q$ is plotted in addition to $m_\pi$.   The point at which
these two curves cross is called the Mott temperature, $T_{M_\pi}$.   For
$T>T_{M_\pi}$, the pion is no longer a bound state, but is a resonance, with 
a finite width that is not shown here.    Similarly we have plotted
$m_q+m_s$, from which the kaonic Mott temperature $T_{M_K}$ is defined.
For $T>> T_{M_K}$, the kaon is also a resonance with a finite 
width

%
%
\begin{center} 
\begin{figure} 
\epsfig{file=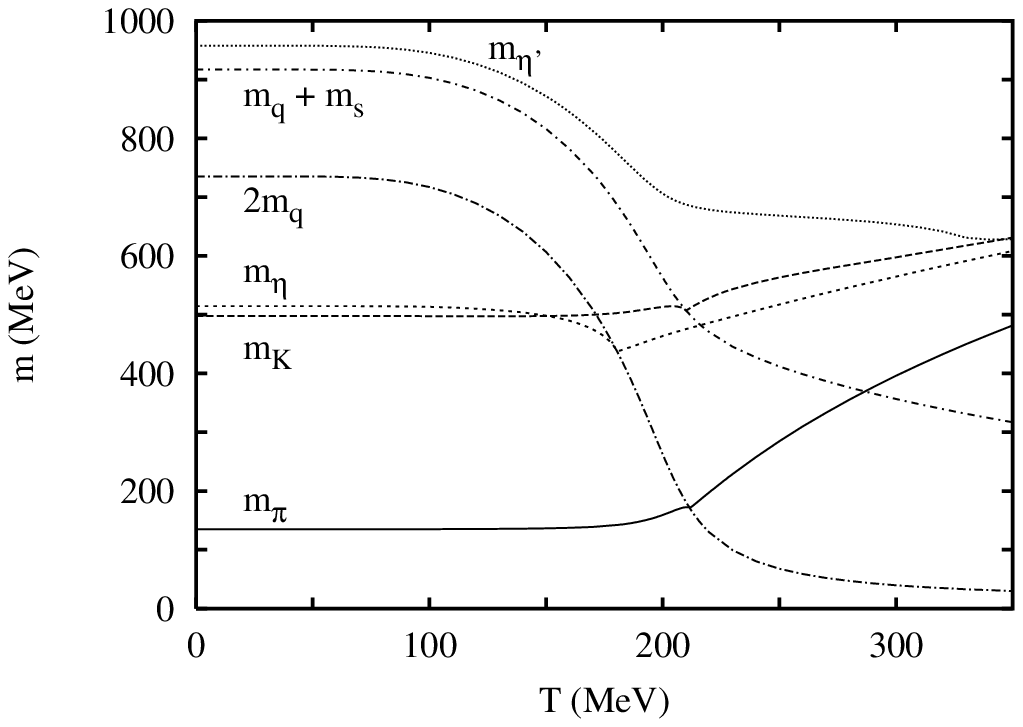,angle=0}
\caption{Temperature dependence of the pseudoscalar meson masses,
as well as that of $2m_q$ and $m_q+m_s$.  
\cite{hadron}.}
\end{figure}   
\end{center}   
 
%
%
\begin{center} 
\begin{figure} 
\epsfig{file=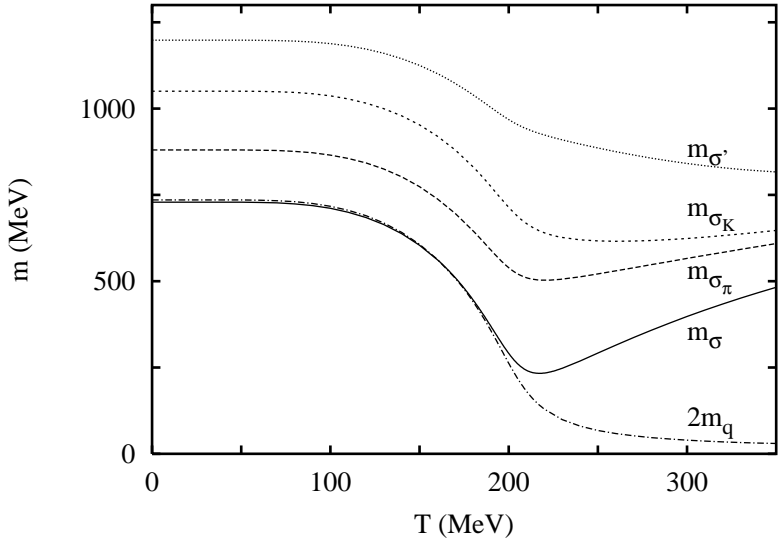,angle=0}
\caption{Temperature dependence of the scalar meson masses
and $2m_q$
\cite{hadron}.}
\end{figure}   
\end{center}

These graphs deserve some comment.    Firstly let us compare them with 
the figure showing the meson masses obtained via lattice gauge theory,
Fig.~2.   We note first that there are some fundamental differences in
obtaining these graphs: (a) Figs.~8 and 9 show so-called {\it pole}
masses, while Fig.~2 gives {\it screening}\  masses.    Nevertheless, it has
been shown that, in the NJL model, the temperature behavior of
screening masses and pole masses is {\it qualitatively similar} 
\cite{flor}, although
quantitatively somewhat different.   Since we cannot hope for any 
quantitative agreement at this stage, it is justifiable to make a
comparison.
(b) The NJL model calculation shown is for SU(3), while the lattice
calculation is SU(2).    With these points in mind, note that the
$\sigma$ and $\pi$ mesons from figures 8 and 9 become degenerate at 
high temperatures, as observed also in Fig.~2.    However, there is no
undershooting of the $\sigma$ meson.   The meson labelled $m_{\sigma_\pi}$
of Fig.~8 corresponds to the $a_0$ of Fig.~2.   Here one observes
qualitatively the same behavior, i.e. that this scalar meson also decreases
strongly in the phase transition region.   Thus one has an aesthetically
pleasing agreement between the NJL model and the results obtained by
lattice gauge theory for meson masses at this level.

A direct comparison of Fig.~7 with the results of chiral perturbation theory, 
i.e. with 
Fig.~5 is problematic.   We simply make some comments: (a) the physics
underlying Figs.~5 and 7 is completely different.    Fig.~5 is obtained by
constructing meson loops (the mesons are regarded as structureless
point-like objects), while in Fig.~7, the pion is constructed from a
quark-antiquark loop.   Meson loops form corrections to this calculation and
would be of the next order in $1/N_c$.    Such corrections have
 in fact been
evaluated, and it has been found that the leading order $T^2$
dependence of Eq.(\ref{e:mpit}) is recovered \cite{flor2}.
    The fact that in the final
analysis the curve of $M_\pi$ as a function of temperature is finally {\it
decreasing} for chiral perturbation theory in Fig.~5,
 is in strong contradiction to both Figs.~2 and 8.

\subsection{Bulk thermodynamic quantities}

In the last subsection, we have indicated the successes of the NJL model
in calculating the order parameter and masses as a function of 
temperature.   In this subsection, we turn to bulk thermodynamic quantities.
Here we will see that the model does not do as well, and that the lack of
confinement makes itself strongly evident at low temperatures, while 
the cutoff of the model is a hindrance at high temperatures.   We start
with the thermodynamical potential $\Omega$, calculated in the grand 
canonical ensemble.   Given an interaction between fermions that is 
4-point in nature such as in Eq.(\ref{e:lnjlsu2}), $\Omega$ can be
calculated quite generally as \cite{reviews,fetter}
\begin{equation}
\Omega =\Omega_0+ \int_0^1\frac{d\lambda}{\lambda}\frac12\int
\frac{d^3p}{(2\pi)^3}\frac 1\beta\sum_n \exp(i\nu_n\eta){\rm Tr}[\Sigma^
\lambda(\nu_n,\vec p) S^\lambda(\nu_n,\vec p)],
\label{e:omega1}
\end{equation}
where $\Omega_0$ is the thermodynamic potential in the absence of
interactions, and $\sum^\lambda$ and $S^\lambda$ designate the Matsubara
self-energy and Green function associated with the system.   The superscript
$\lambda$ refers to the fact that both $S$ and $\Sigma$ are to be evaluated
with the introduction of an artifical coupling that multiplies the 
interaction Lagrangian ${\cal L}_{int}$.    The Matsubara frequencies for
fermions are, as required, odd, i.e. $\nu_n = (2n+1)\pi/\beta$, with
$n=0,\pm1,\pm2\pm3\dots$.

For the NJL Lagrangian of Eq.(\ref{e:lnjlsu2}), in the mean field
approximation, it is not necessary to apply Eq.(\ref{e:omega1}).   A
straightforward calculation gives
\begin{equation}
\Omega=\Omega_q = - 2N_cN_f\int\frac{d^3p}{(2\pi)^3} E_p - 
\frac{2N_cN_f}\beta\int\frac{d^3p}{(2\pi)^3} \ln[1+e^{-\beta(E_p+\mu)}]
[1+e^{-\beta(E_p-\mu)}],
\label{e:omegaq}
\end{equation}
with $E^2_p=\vec p^2+m^2$.  As is evident from the label $q$, this appears
to
be a thermodynamic potential generated solely by the {\it quark} degrees of 
freedom.

We note also that the thermodyamical properties can only be measured
relative to the physical vacuum,
\begin{equation}
\Omega^{phys}_{vac} = \Omega(T=0,\mu=0,m(0,0)),
\label{e:omegaphys}
\end{equation}
which, for the mean field approximation, corresponds to
\begin{equation}
(\Omega^{phys}_{vac})_{mf} = \frac {(m-m_0)}{4G} - 2N_cN_f\int\frac
{d^3p}{(2\pi)^3} E_p.
\label{e:omphysmf}
\end{equation}

To introduce {\it mesonic} \ degrees of freedom, it is necessary to go
beyond the mean field approximation to include the next set of terms in the
$1/N_c$ expansion.   The self-energy in this case includes effective 
interactions in both the scalar and pseudoscalar channels \cite{pengfei},
\begin{eqnarray}
\Sigma_{fl}^\lambda(\nu_n,\vec p) = &-&\frac 1\beta\sum_{n'}\int\frac{d^3q}
{(2\pi)^3}[S(\nu_{n'},\vec q)V_\sigma^\lambda(\nu_n- \nu_{n'},\vec p - \vec
q )\nonumber \\
&+&i\gamma_5\vec \tau S(\nu_{n'},\vec q) i\gamma_5\vec \tau V_\pi^\lambda
(\nu_n-\nu_{n'},\vec p-\vec q)],
\label{e:star}
\end{eqnarray}
and is constructed on summing the Fock and infinite RPA series that 
contribute to the self-energy in this order.   Here
\begin{equation}
V_M^\lambda(\omega,\vec q) = -2G\lambda [1-2G\lambda\Pi_M(\omega,\vec
q)]^{-1},
\label{e:v}
\end{equation}
and the irreducible polarization in the mesonic channel  
\begin{equation}
\Pi_M(\omega_m,\vec p) = \int\frac{d^3q}{(2\pi)^3}\frac 1\beta\sum_n
{\rm Tr} \Gamma_M S(\nu_n+\omega_m,\vec q + \vec p) \Gamma_M S(\nu_n,\vec
q),
\label{e:pol}
\end{equation}
is determined by the vertex $\Gamma_M$ for that channel.   Inserting
Eq.(\ref{e:star}) into Eq.(\ref{e:omega1}) yields the fluctuating part of
the thermodynamic potential,
\begin{equation}
\Omega_{fl} = \sum_m\frac {N_M}2\int\frac{d^3p}{(2\pi)^3}\frac 1\beta
\sum_n e^{i\omega_n\eta} \ln[1-2G\Pi_M(\omega_n,\vec p)].
\label{e:omegafl}
\end{equation}
The nature of this term is revealed on performing the frequency sum.
One has
\begin{equation}
\Omega_{fl} = \Omega_\pi + \Omega_\sigma,
\label{e:decomp}
\end{equation}
where, for each species $M=\pi$ or $\sigma$,
\begin{eqnarray}
\Omega_M = &-&N_M\int\frac{d^3p}{(2\pi)^3}\int_0^\infty d\omega [\frac12
\omega + \frac 1\beta\ln(1-e^{-\beta\omega})]\frac 1{2\pi i} \nonumber\\
&& \times\frac d{d\omega} \ln\frac{1-2G\Pi_M(\omega+i\epsilon,\vec p)}
{1-2G\Pi_M(\omega-i\epsilon,\vec p)}.
\label{e:trouble}
\end{eqnarray}
Some analysis shows that a simple approximation for the polarization near
the pole, {\it i.e.}
\begin{equation}
1-2G\Pi_M(\omega,\vec p) = (\omega^2 - E_M^2) \times {\rm const}
\label{e:poleapp}
\end{equation}
leads to 
\begin{equation}
\Omega_M=N_M\int\frac{d^3p}{(2\pi)^3} [\frac 12 E_M + \frac 1\beta
\ln(1-e^{-\beta E_M})],
\label{e:expect}
\end{equation}
exactly as one would expect for the thermodynamic potential given 
bosonic degrees of freedom.   The pole
approximation is however insufficient, as one integrates over all energies,
and in practice, the fact that the bound states also become delocalized
resonances at the Mott point must also be accounted for.    This has been done
in introducing phase shifts in each channel \cite{pengfei}.

In order to calculate the pressure, the physical vacuum given by 
 Eq.(\ref{e:omegaphys}) 
must be reevaluated  to include
a term from $\Omega_{fl}$.    One now  has
\begin{equation}
\Omega^{phys}_{vac} = (\Omega^{phys}_{vac})_{mf} + (\Omega^{phys}_{vac})_{
fl}.
\end{equation}
and the pressure density is now
\begin{equation}
p=-\Omega_q-\Omega_{fl} + \Omega^{phys}_{vac}.
\end{equation}
In Figs.~10 and 11,  the pressure and associated energy  densities evaluated
from this thermodynamic potential are shown.   

%
%
\begin{center} 
\begin{figure} 
\epsfig{file=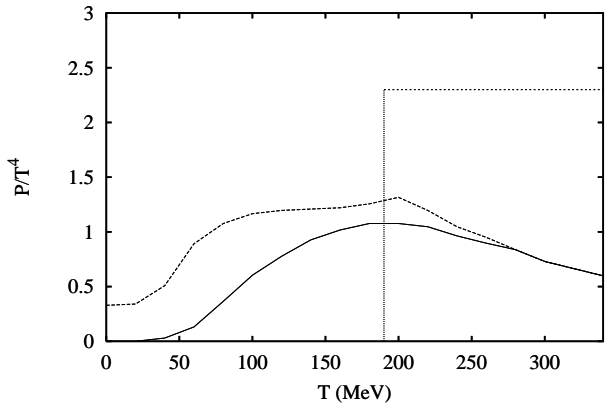,angle=0,scale=0.4}
\caption{The pressure density, scaled by $T^4$, is shown as a function of
temperature.    The lower curve is for the mean field case only, the upper
includes fluctuations (mesons).    The vertical line indicates the critical
temperature and the horizontal one the Stefan-Boltzmann limit for an ideal
quark gas
\cite{pengfei}.}
\end{figure}   
\end{center}

In Fig.~10, one sees that the lower curve, corresponding to the mean field
approximation calculation represents the quark degrees of freedom.   There
is an appreciable pressure that arises from this term,
{\it i.e.}\ from the quark degrees of freedom, for temperatures
$T<T_c$, which is indicated by the vertical line.    Including mesonic 
degrees of freedom rectifies the behavior at small temperatures, but still
leaves a large intermediate range of temperatures $T<T_c$ that
is dominated by these 
unphysical quark degrees of freedom.  This is thus a direct
consequence of the missing feature of confinement.   The sharp rise in the
pressure density shown in Fig.~3 cannot be modelled by a non-confining
theory.    At high temperatures, $T>T_c$, there is a small contribution from
the mesonic degrees of freedom, that exist as correlated states with a
finite width in the plasma.   The main contribution arises 
however here from the quark 
degrees of freedom.   The actual value obtained for the pressure density
underestimates the Stefan-Boltzmann
limit (shown as a horizontal line), since there is a finite cutoff on the
quark momenta.   Relaxing this constraint would lead to the pressure density
approaching a constant.

%
%
\begin{center} 
\begin{figure} 
\epsfig{file=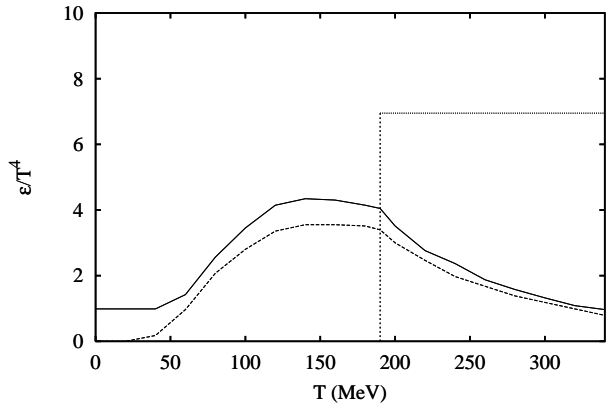,angle=0,scale=0.4}
\caption{The energy density, scaled by $T^4$, is shown as a function of
temperature.    The lower curve is for the mean field case only, the upper
includes fluctuations (mesons).    The vertical line indicates the critical
temperature and the horizontal one the Stefan-Boltzmann limit for an ideal
quark gas
\cite{pengfei}.}
\end{figure}   
\end{center}

Similar comments can be made for the energy density:   the intermediate
temperature range $50$MeV$<T<T_c$ is dominated by quark degrees of freedom,
indicating the lack of confinement.   The high temperature values $T>T_c$
do not approach the Stefan-Boltzmann limit, due to the cutoff.   

In concluding this subsection, one sees that one needs to include
confinement
in some fashion in order to be able to regain the lattice picture.   From
a thermodynamic point of view, the high temperature regime about $T\simeq 
T_c$ is probably best described by the model, in the sense that only quark
degrees of freedom plus correlated mesonic states are present.   In the
next subsection, we thus study elastic quark-antiquark scattering about this
point and indicate that a divergence occurs in the cross-section at $T=T_M$
 and that the phenomenon of critical scattering as a consequence of the 
chiral phase transition is observed.

\subsection{Critical opalescence in the quark-antiquark channel}

In this section, we examine the behavior of the quark-antiquark
scattering amplitude in the NJL model in the vicinity of the Mott
temperature, which replaces the critical temperature when finite
current quark masses are used.   In SU(3), there are seven
independent processes out of a total of fifteen for quark-antiquark
scattering, taking isospin and charge conjugation symmetry into
account.   These are listed in Table~3.   Mesons that can be 
exchanged in the $s$ and $t$ channels, as are given by the
Feynman diagrams of Fig.~12 are also listed.

\begin{table}[ht]
\begin{center}
\begin{tabular}{c l l } \hline
Process &  Exchanged mesons ($s$ channel) &  Exchanged mesons
($t$ channel) \\ \hline
$u\bar d\rightarrow u\bar d$ & $\pi$, $\sigma_\pi$ &
$\pi$, $\eta$, $\eta'$, $\sigma_\pi$, $\sigma$, $\sigma'$ \\
$u\bar s \rightarrow u\bar s$ & $K$, $\sigma_K$ &
$\eta$, $\eta'$, $\sigma$, $\sigma'$ \\
$u\bar u \rightarrow u\bar u $ & $\pi$, $\eta$, $\eta'$, $\sigma_\pi$,
$\sigma$, $\sigma'$ &
$\pi$, $\eta$, $\eta'$, $\sigma_\pi$, $\sigma$, $\sigma'$ \\
$u\bar u\rightarrow d\bar d$ & $\pi$, $\eta$, $\eta'$, $\sigma_\pi$,
$\sigma$, $\sigma'$  & $\pi$, $\sigma_\pi$ \\
$u\bar u\rightarrow s\bar s$ & $\eta$, $\eta'$,  $\sigma$, $\sigma'$ 
& $K$, $\sigma_K$ \\
$s\bar s \rightarrow u\bar u$ & $\eta$, $\eta'$,  $\sigma$, $\sigma'$
&  $K$, $\sigma_K$ \\
$s\bar s\rightarrow s\bar s$ &  $\eta$, $\eta'$,  $\sigma$, $\sigma'$
&  $K$, $\sigma_K$ \\ \hline 
\end{tabular}
\end{center}
\caption{Independent processes for $q\bar q$ scattering.}
\end{table}

%
%
\begin{center}
\begin{figure}
\epsfig{file=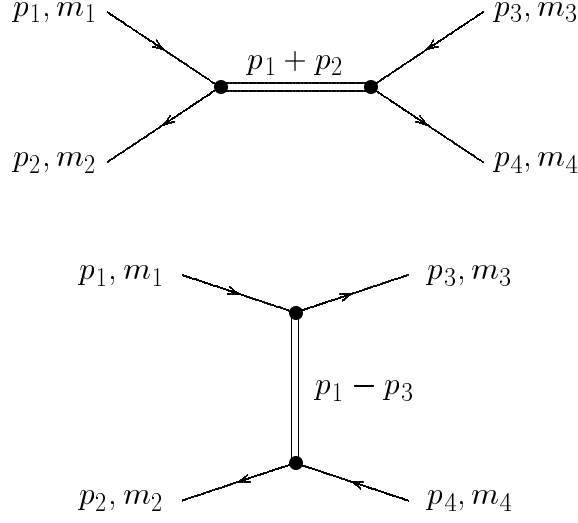,angle=0,scale=0.5}
\caption{Feynman diagrams for elastic $q\bar q$ scattering 
within the NJl model in an expansion to lowest order
in $1/N_c$.}
\end{figure}
\end{center}

The transition amplitudes can be written as
\begin{eqnarray}
-i{\cal M}_s &=& \delta_{c_1,c_2}\delta_{c_3,c_4} \bar v(p_2)Tu(p_1)
[i{\cal D}_s^S(p_1+p_2)]\bar u(p_3) T v(p_4) \nonumber \\
&& + \delta_{c_1,c_2}\delta_{c_3,c_4} \bar v(p_2)(i\gamma_5T)u(p_1)[i{\cal 
D}_s^P(p_1+p_2)]\bar u(p_3)(i\gamma_5T)v(p_4), \nonumber \\
\label{e:ms}
\end{eqnarray}
and
\begin{eqnarray}
-iM_t &=& \delta_{c_1,c_3}\delta_{c_2,c_4}\bar u(p_3) T u(p_1)[i{\cal D}_t^S
(p_1-p_3)] \bar v(p_2)T v(p_4) \nonumber \\
&& \delta_{c_1,c_3}\delta_{c_2,c_4}\bar u(p_3) (i\gamma_5T)u(p_1) [i{\cal
D}_t^P(p_1-p_3)]\bar v(p_2) (i\gamma_5 T) v(p_4), \nonumber \\
\end{eqnarray}
where $T$ selects the isospin eigenvalue for a particular channel, and 
$D_{s,t}^{S,P}$ is the $s$ or $t$ channel, scalar or pseudoscalar 
quark-antiquark scattering amplitude, and which can be constructed from the
corresponding polarization function.   It has a simple form, for example 
\cite{spk}
\begin{equation}
{\cal D}_\pi(p_0,\vec p) = \frac {2G^{eff}}{1-2G^{eff}\Pi^P_{u\bar u}(
p_0,\vec p)},
\end{equation}
where $G^{eff}$ is an effective SU(3) coupling strength in the pionic
channel \cite{spk}.

The differential cross section is constructed in the usual fashion as
\begin{equation}
\frac{d\sigma}{dt} = \frac1
{16\pi[s-(m_u+m_s)^2][s-(m_u-m_s)^2] }\frac 1{4N_c^2} \sum_{s,c}
|{\cal M}_s-{\cal M}_t|^2,
\end{equation}
while the total cross section is evaluated as
\begin{equation}
\sigma = \int \ dt\  \frac{d\sigma}{dt} [1-f_F(\beta E_3)][1-f_F(\beta
E_4)],
\end{equation}
introducing a Fermi blocking factor for the final states.    Here $E_i^2=
p_i^2 + m_i^2$, where $i=3,4$.

In Fig.~13, we show the total cross section for light quarks in the initial
state, as a function of $\sqrt{s}$, at a temperature $T=215$ MeV, which lies
slightly higher than the pion and kaonic Mott temperatures, $T_{M_\pi} =
212$MeV and $T_{M_K}=210$MeV.    Both pions and kaons are sharp resonances
now.   At higher values of the temperature, these become broader resonances
in the cross-section, as shown in Fig.~14.   At the Mott temperature itself,
when quarks bind into hadrons, the intermediate states in the $s$ channel
give rise to \emph{infinite} cross sections at threshold.   This feature,
which also appears in other processes like $\pi\pi\rightarrow \pi\pi$
\cite{kalinov}, $\pi\gamma\rightarrow\pi\gamma$ \cite{dorokh} or $q\bar q
\rightarrow \gamma\gamma$ \cite{spk2,blaschke} is akin to the phenomenon of
critical opalescence.  This  has been discussed in some detail in
Ref.\cite{mothballs}, and the interested reader is referred to this.

%
%
\begin{center}
\begin{figure}
\epsfig{file=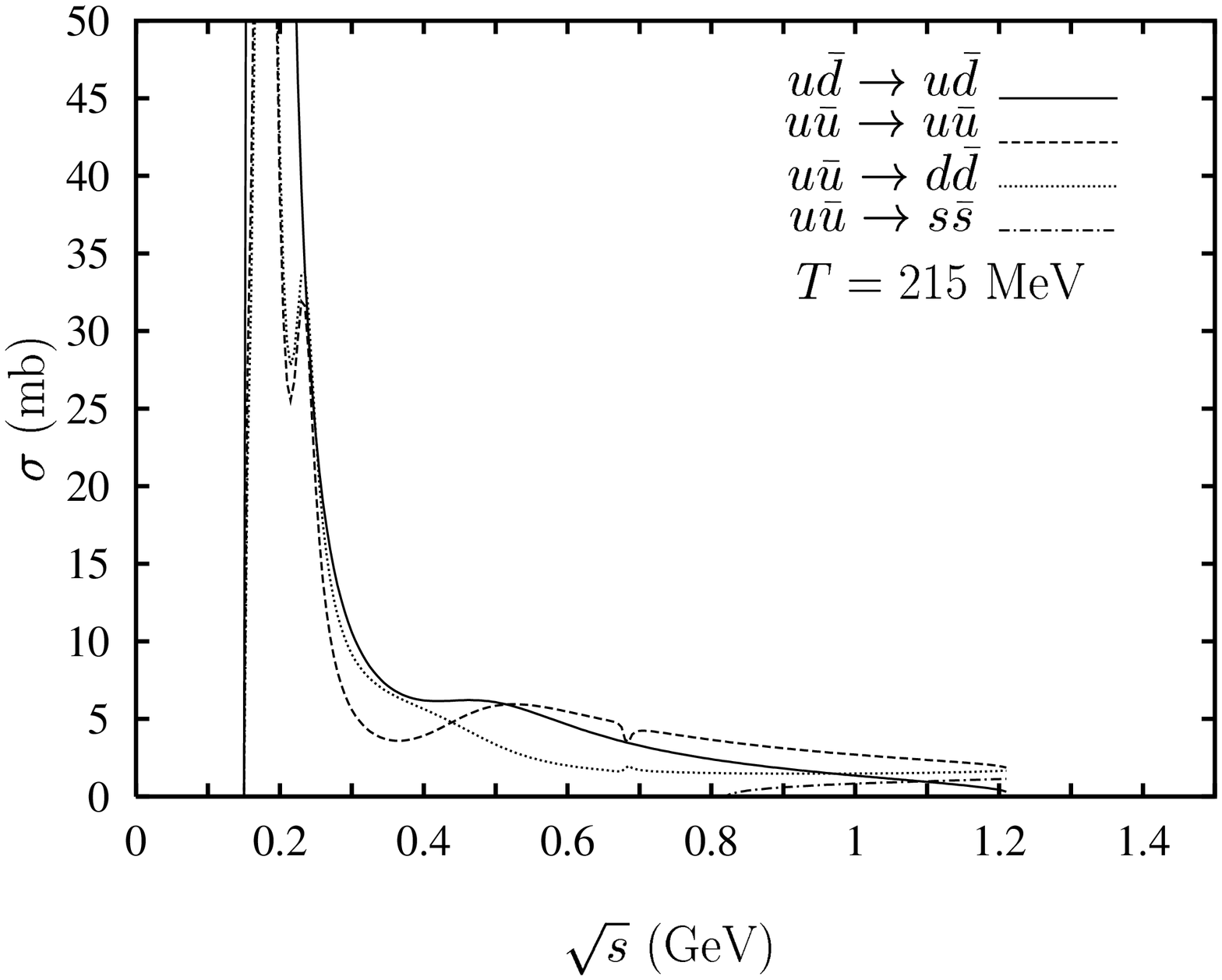,angle=-90,scale=0.5}
\caption{Total cross section for $q\bar q$ scattering with only light
quarks in the initial state, shown as a function of $\sqrt s$, at $T=215$
MeV.}
\end{figure}
\end{center}

%
%
\begin{center}                             
\begin{figure}                             
\epsfig{file=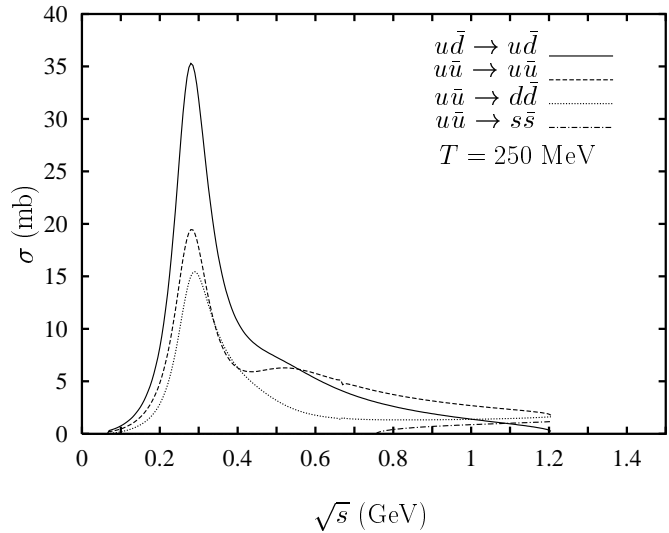,angle=-90,scale=0.425}
\caption{Total cross section for $q\bar q$ scattering with only light
quarks in the initial state, shown as a function of $\sqrt s$, at $T=250$
MeV.}                                      
\end{figure}                               
\end{center}

\section{Non-equilibrium formulation and transport equation}

The considerations of the first two sections discussed properties
of chiral systems {\it in equilibrium}.
   If it were possible to measure any of
the associated changes at the phase transition temperature,
 there would be no need for
further discussion.   However, because of the nature of confinement, we are 
unable to observe critical scattering directly, nor any of the other
dramatic changes in pion properties.     One tool for examining quark
matter is via heavy ion collisions, and as such, over the short time
scales over which collisions occur, it is unclear whether both thermal
and chemical equilibrium can be reached during a collision.   For this
reason, we wish to investigate what the effects are of a condensate that
changes with the medium, as well as medium dependent cross sections in a 
non equilibrium scenario.  

There are several formal aspects that have to be understood before one can
attempt actual collision simulations.    Firstly one can set up an exact
formal description of a relativistic fermionic system that is out of
equilibrium via the method of Schwinger and Keldysh.   From a 
heuristic point of view, however, we have a good understanding of the
classical Boltzmann equation, so that it is important to establish a link
between the two from which one can then go further.    In doing so, one
generally has a field theory with retarded and advanced Green functions.
However, if we examine the collision term of the Boltzmann equation, we see
that we require cross sections.  However,  we only know how to calculate these
using causal Green functions.    So we have to find a link telling us which
level of approximation requires which Feynman graphs.

The content of this lecture is summarized briefly in the next paragraphs.
(a)  We wish to start from a chosen Lagrangian that gives a microscopically
correct description of the world, and to formulate a non-equilibrium
theory via a matrix of Green functions $S^{ij}(x_1,x_2)$ ($i$ and $j$ will
be defined later!).     This matrix of Green functions 
satisfies a matrix form of the 
Schwinger-Dyson equations, which as usual, cannot be solved exactly.
(b) Some technical aid is required at this point.   A centre of mass
variable $X=(x_1+x_2)/2$ and relative coordinate $u=x_1-x_2$ are introduced,
and one {\it Wigner transforms}\ the matrix of Green functions.   This is
simply a Fourier transform with respect to the relative coordinate $u$.   At
this point, the equations are still exact.   (c)  Now one seeks 
methods of solution.   For
a fermionic system, the exact method would involve making a spinor
decomposition of the Green functions, and we would have 32 coupled equations
to solve!  This is simply too difficult, in particular for an expanding
system, for which spatial gradients are important, and so we turn rather to
making the {\it quasiparticle assumption},\ which, coupled with an expansion
in powers of $\hbar$, leads to the well-known kinetic theory of Boltzmann,
here in relativistic form.

All that has been discussed is quite general for any fermionic theory. 
Using the Lagrange density of the Nambu--Jona-Lasinio Lagrangian with an
expansion in $1/N_c$ illustrates how extensions to the standard binary
collision forms in the
 Boltzmann equation come about, and clears the issue of the 
content of Feynman graphs for the cross sections that occur in the 
Boltzmann equation.

\subsection{Closed time path -- Schwinger-Keldysh formalism}

There are several excellent texts that exist that cover the basics of 
the Schwinger-Keldysh formalism \cite{schwinger,keldysh} for Green functions
not in equilibrium.    Detailed reviews using path integrals can be found
in \cite{chou}, while the more standard operator approach is to be found
in \cite{malfliet,degroot,landau}.   Most confusing in this subject is
simply notation:  All the listed references use different ones.    I shall
conform to that of Landau\footnote{This differs from the labelling of
\cite{malfliet} by a minus sign.   Off-diagonal self-energies also differ by
a minus sign.}, which is particularly transparent in setting up rules for
a perturbative diagrammatic expansion.    

Central to the problem of non-equilibrium systems is that the description
via
a single causal Green function alone, is inadequate.   One requires the
four Green functions, 
\begin{eqnarray}
i\hbar S^c(x,y) &=& \left <T\psi(x) \bar \psi(y)\right > = i\hbar
S^{--}(x,y)\nonumber
\\
i\hbar S^a(x,y) &=& \left <\tilde T\psi(x)\bar\psi(y)\right > = i\hbar
S^{++}(x,y)
\nonumber \\  
i\hbar S^{>}(x,y) &=& \left <\psi(x)\bar\psi(y)\right > = i\hbar S^{+-}(x,y)
\nonumber
\\
i\hbar S^<(x,y) &=& -\left <\bar\psi(y)\psi(x)\right > = i\hbar S^{-+}(x,y),
\label{gfunctions}
\end{eqnarray}
{\it i.e.}\ the causal and  acausal propagators $S^c$ and $S^a$,
 $S^>$ and $S^<$.   In Eq.(\ref{gfunctions}),  $T$ is the standard time
ordering operator,
\begin{equation}
T(O(x)O(y)) = \theta(x_0-y_0) O(x)O(y) - \theta(y_0-x_0) O(y)O(x),
\end{equation}
and $\tilde T$ the {\it anti}time ordering operator,
\begin{equation}
\tilde T(O(x)O(y)) = \theta(y_0-x_0) O(x)O(y) -\theta(x_0-y_0)  O(y)O(x).
\end{equation}
On the right hand side of Eq.(\ref{gfunctions}), the superscripts $ij=+,-$ 
have been introduced (these were mentioned in the introduction to this
section).   This is
an arbitrary but useful convention for constructing a matrix notation for
summarizing the Green functions,
 \begin{equation}\label{green}
    \underline{S} = \left( \begin{array}{cc}
                              S^{--} & S^{-+} \\
                              S^{+-} & S^{++}
                           \end{array} \right).
  \end{equation}
It  is automatically achieved by introducing the closed time path of
Fig.~15, and setting the fields that occur in the Green function $S^{ij}$
 on the 
$i$th or $j$th branch respectively.

%
%
\begin{center}                             
\begin{figure} 
\epsfig{file=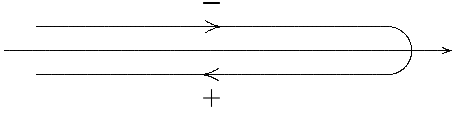,angle=0,scale=0.425}
\caption{Closed time path on which the Green functions are defined}
\end{figure}                  
\end{center}

There are many interlinking relationships that follow simply from the 
definition of the Green functions.   For example, $S^c$ and $S^a$
are related to $S^>$ and $S^<$ via
\begin{eqnarray}
S^{--}(x,y) &=& \theta(x_0-y_0) S^{+-}(x,y) + \theta(y_0-x_0) S^{-+}(x,y) 
\nonumber \\
S^{++}(x,y) &=& \theta(y_0-x_0) S^{+-}(x,y) + \theta(x_0-y_0) S^{-+}(x,y).
\end{eqnarray}
All four Green functions are not independent, since
\begin{eqnarray}
S^{--}(x,y) + S^{++}(x,y) &=& S^{-+}(x,y) + S^{+-}(x,y) \nonumber \\ 
                          &=& S^K(x,y) = -\frac i\hbar\langle[\psi(x),
\bar\psi(y)]\rangle,
\end{eqnarray}
defining the Keldysh Green function.   In addition, one can define the
retarded and advanced Green functions
\begin{eqnarray}
i\hbar S^R(x,y) &=& \theta(x_0-y_0) \langle\{ \psi(x),\bar\psi(y)\}\rangle 
\nonumber \\
i\hbar S^A(x,y) &=& -\theta(y_0-x_0) \langle\{\psi(x),\bar\psi(y)\}\rangle,
\end{eqnarray}
which are also related to the $S^{ij}$ via
\begin{eqnarray}
S^R(x,y) &=& S^{--}(x,y) - S^{-+}(x,y) = S^{+-}(x,y) - S^{++}(x,y) \nonumber
\\
S^A(x,y) &=& S^{--}(x,y) - S^{-+}(x,y) = S^{-+}(x,y) - S^{++}(x,y),
\label{e:relati} 
\end{eqnarray}
which can also be verified directly from the definitions of these functions.
One could consider working with the matrix of independent functions
\begin{equation}
    \underline{S}' = \left( \begin{array}{cc}
                                 0       & S^{\rm A} \\
                               S^{\rm R} & S^{\rm K}
                            \end{array} \right)  ,
  \end{equation}
but I will not do so in this chapter.  Nevertheless, the retarded and 
advanced Green functions play a special role.   Due to their simple analytic
structure, plus the fact that the equations of motion that they satisfy 
(see Eq.(\ref{ret}) later!) are closed, means that one usually can find a simple
analytic form for these functions.

The matrix of self-energies is defined now via the Dyson equation,
 \begin{eqnarray}
    \underline{S}(x,y) &=& \underline{S^{0}}(x,y)
                  + \int d^{4}z d^{4}w \underline{S^{0}}(x,w)
                                        \underline{\Sigma}(w,z)
                                        \underline{S}(z,y)\nonumber \\
                    &=& \underline {S^{0}}(x,y)
                  + \int d^{4}z d^{4}w \underline{S}(x,w)
                                        \underline{\Sigma}(w,z)
                                        \underline{S^0}(z,y).
\label{dyson}
  \end{eqnarray}
Pictorially, one can for example examine one element of this equation --
say $S^{++}$.   The equation that this function satisfies is given in
Fig.~16, using an obvious notation.    Thus one sees that all components of 
the self-energy are in fact required in order to evaluate one single component
of $\underline{S}$.

%
%
\begin{center}                             
\begin{figure} 
\epsfig{file=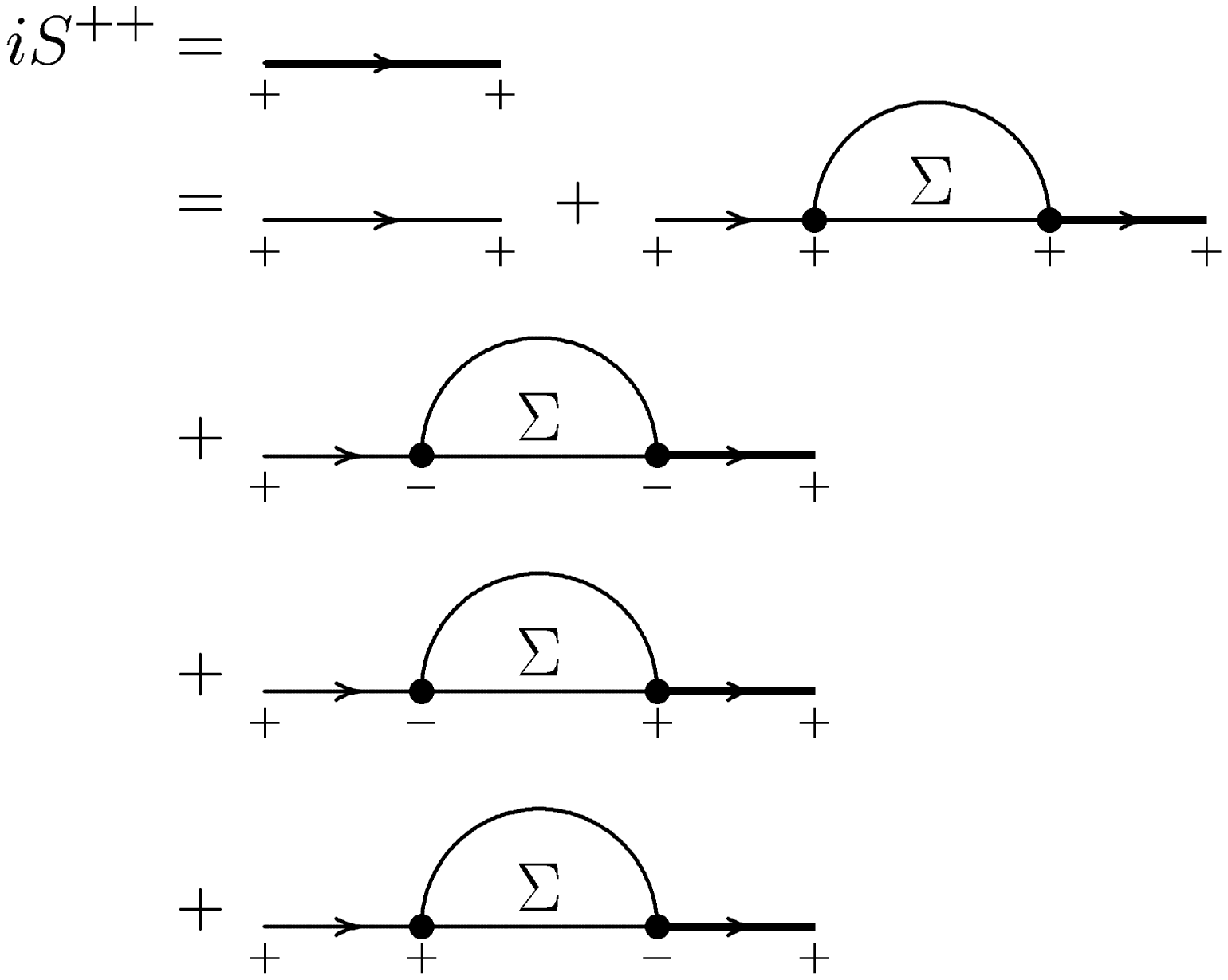,angle=0,scale=0.425}
\caption{Dyson equation for one component of the matrix of Green functions.}
\end{figure}                  
\end{center}

From the Dyson equation, one can derive the equations of motion for the
components of ${\underline{S}}$, which are summarized as
  \begin{equation}
    \left( i \hbar \not\!\partial_{x} -m_0 \right) \underline{S}(x,y) =
        \underline{ \sigma_{z} } \delta^{4} (x-y)
                  + \int d^{4}z  \underline{ \sigma_{z} }
                                        \underline{\Sigma}(x,z)
                                        \underline{S}(z,y) ,
 \label{eq:401}
 \end{equation}
where
  \begin{equation}
    \underline{ \sigma_{z} } = \left( \begin{array}{cc}
                           1 &  0 \\
                           0 & -1
                        \end{array}\right).
  \end{equation}
By defining the retarded and advanced self-energies as
\begin{eqnarray}
\Sigma^R &=& \Sigma^{--} + \Sigma^{-+} \nonumber \\
\Sigma^A &=& \Sigma^{--} + \Sigma^{+-}, 
\end{eqnarray}
one finds the corresponding Dyson equations for $\underline{S}'$,
\begin{equation}
   \underline{S'}(x,y) = \underline{S'}^{(0)}(x,y)
                  + \int d^{4}z d^{4}w \underline{S'}^{(0)}(x,w)  
                                        \underline{\Sigma'}(w,z)
                                        \underline{S'}(z,y),
\label{a16}
  \end{equation}
from which one sees that the Dyson equations for $S^R$ and $S^A$ are 
individually closed,
\begin{equation}
    S^{\rm R,A}_{\beta \gamma}(x,y) = S^{\rm R,A(0)}_{\beta \gamma}(x,y)
                  + \int d^{4}z d^{4}w S^{\rm R,A(0)}_{\beta\mu}(x,w)
                                       \Sigma^{\rm R,A}_{\mu \nu}(w,z)
                                       S^{\rm R,A}_{\nu \gamma}(z,y)
  \end{equation}
with corresponding equation of motion
 \begin{equation}
    \left( i \hbar \not\!\partial_{x} -m_0 \right)_{\alpha\beta} S^{\rm
R,A}_
{\beta\gamma}(x,y) =
         \delta_{\alpha \gamma} \delta^{4} (x-y)
                  + \int d^{4}z  \Sigma^{\rm R,A}_{\alpha \beta}(x,z)
                                 S^{\rm R,A}_{\beta \gamma}(z,y),
\label{ret}        
  \end{equation}   
while the equation for the Keldysh function is integrodifferential,
\begin{equation}   
 \left( i \hbar \not\!\partial_{x} -m_0 \right) S^K =
\int(\Sigma^K S^A + \Sigma^R S^K) d^4z.
\end{equation}     
For a free particle, it is useful to note that 
the solution for the retarded and advanced functions
follows immediately as
\begin{equation}   
S^{R,A}(p) = \frac{\not p + m_0}{p^2 - m_0^2 \pm i\epsilon p_0}
.                  
\label{sretfree}   
\end{equation}     

\subsection{Transport and constraint equations}

Of the matrix of Green functions, consider only the equation of motion for
$S^{-+}(x,y)$ that follows from Eq.(\ref{eq:401}).   This is
\begin{equation}            
 \left( i \hbar \not\!\partial_{x} - m_0 \right)_{\alpha \beta}
              S^{-+}_{\beta \gamma}(x,y)     =\int d^{4} z \left\{
\Sigma^{--}_{\alpha \beta} ( x,z )
                               S^{-+}_{\beta \gamma}(z,y)
                            + \Sigma^{-+}_{\alpha \beta} ( x,z )
                               S^{++}_{\beta \gamma}(z,y) \right\}
\label{s}                   
\end{equation}              
In a similar fashion, one can derive the equation of motion
  \begin{eqnarray}            
&&S^{-+}_{\alpha \beta}(x,y)  
           \left( -i \hbar \overleftarrow{\not\!\partial_{y}}
                                 -m_0 \right)_{\beta \gamma}  =
 \int d^{4}z                
        \{ - S^{--}_{\alpha \beta}(x,z)
                  \Sigma^{-+}_{\beta \gamma}(z,y) \nonumber \\
&& \quad\quad \quad                - S^{-+}_{\alpha \beta}(x,z)
                  \Sigma^{++}_{\beta \gamma}(z,y) \}.
\label{scross}              
\end{eqnarray}        
It turns out to be slightly more convenient to cast these equations in an 
alternative form, using the relations Eq.(\ref{e:relati}) between the Green
functions, and similar ones for the self-energies.   We write
  \begin{eqnarray}          
     \left( i \hbar \not\!\partial_{x} - m_0 \right)_{\alpha \beta}
              S^{-+}_{\beta \gamma}(x,y)
&=&\int d^{4} z \{ \Sigma^{-+}_{\alpha \beta} (x,z)
         S^{+-}_{\beta \gamma} (z,y)
      - \Sigma^{+-}_{\alpha \beta} (x,z)
         S^{-+}_{\beta \gamma} (z,y)                    \nonumber \\
    & &  + \Sigma^{\rm A}_{\alpha \beta} (x,z)
            S^{-+}_{\beta \gamma} (z,y)
         - \Sigma^{-+}_{\alpha \beta} (x,z)
            S^{\rm R}_{\beta \gamma} (z,y)  \} \nonumber \\
\label{s1again}             
\end{eqnarray}              
and
\begin{eqnarray}            
     S^{-+}_{\alpha \beta}(x,y)
           \left( -i \hbar \overleftarrow{\not\!\partial_{y}}
                                 -m_0 \right)_{\beta \gamma}
 = \int d^{4}z              
        \{ &-& S^{\rm R}_{\alpha \beta}(x,z)
                  \Sigma^{-+}_{\beta \gamma}(z,y) \nonumber \\
                &+& S^{-+}_{\alpha \beta}(x,z)
                  \Sigma^{\rm A}_{\beta \gamma}(z,y) \},
\label{sagain}              
  \end{eqnarray}            
It is now a tedious technical task to Wigner transform
Eqs.(\ref{s1again}) and (\ref{sagain}).   We illustrate this on a simple
example and then simply give the final result.   Introducing relative and 
centre of mass variables $u=x-y$ and $X=\frac 12(x+y)$, the Wigner transform
of $S(x,y)$ is defined to be
\begin{equation}
S(X,p) = \int d^4 u e^{ip.u/\hbar} S(X+\frac u2,X - \frac u2).
\end{equation}
To Wigner transform say the first term on the left hand side of
Eq.(\ref{s1again}) requires an integral of the form
\begin{eqnarray}
&&\int d^4 u e^{ipu/\hbar}\partial^\mu_y f(x,y) \nonumber \\
&&= \int d^4u e^{ipu/\hbar} (\frac 12 \frac\partial{\partial X_\mu} -
\frac\partial{\partial u_\mu}) f(X+\frac 12u, X-\frac 12u) \nonumber \\
&& = \frac 12 \frac \partial {\partial X_\mu}\int d^4u e^{ipu/\hbar}
f(X+\frac 12 u, X-\frac 12u) \nonumber \\
&& \quad \quad + \int d^4u(\frac \partial{\partial u_\mu}
e^{ipu/\hbar}) f(X+\frac 12 u,X-\frac 12u)\nonumber \\
&& = (\frac 12 \partial_X^\mu + \frac {ip^\mu}\hbar ) f(X,p).
\label{e:explicit}
\end{eqnarray}
Similarly one can show that
\begin{eqnarray}
\partial_x^\mu f(x,y) &\rightarrow& (-i\frac{p^\mu}\hbar
 + \frac 12 \partial^\mu_X)f(X,p)
 \\
f(y)g(x,y)&\rightarrow& f(X)\exp\left(\frac
{i\hbar}2\frac{\overleftarrow\partial}{\partial X^\mu}\frac{\overrightarrow    
\partial}{\partial p_\mu}\right)g(X,p)
 \\
f(x)g(x,y) &\rightarrow& f(X)\exp\left(-\frac
{i\hbar}2\frac{\overleftarrow\partial}{\partial X^\mu}\frac{\overrightarrow
\partial}{\partial p_\mu}\right) g(X,p)
 \\
\int d^4z f(x,z) g(z,y) &\rightarrow& f(X,p)\exp \left(-\frac{i\hbar}2
(\frac{\overleftarrow\partial}{\partial X^\mu}\frac
{\overrightarrow\partial}
{\partial p_\mu} - \frac{\overleftarrow\partial}{\partial p_\mu}
\frac{\overrightarrow \partial}{\partial X^\mu})\right) g(X,p) \nonumber \\
\end{eqnarray}
need be made on Wigner transforming the product functions on the left 
hand side of the last equations.
Applying these relations to Eqs.(\ref{s1again}) and (\ref{sagain}) leads
to the rather complex forms for the equations of motion,
\begin{eqnarray}
\{ i\hbar\gamma^\mu(\frac 12\frac{\partial}{\partial X^\mu}&-& \frac
{ip_\mu}\hbar) -m_0\} S^{-+}(X,p) = \nonumber \\
& &\Sigma^{-+}(X,p) \hat \Lambda
S^{+-}(X,p) - \Sigma^{+-}(X,p)\hat\Lambda S^{-+}(X,p) \nonumber \\
&+& \Sigma^{A}(X,p)\hat \Lambda S^{-+}(X,p) - \Sigma^{-+}
(X,p)\hat\Lambda S^R(X,p) \nonumber \\
\end{eqnarray}
and
\begin{eqnarray}
S^{-+}(X,p)\{-i\hbar\gamma^\mu(\frac12\frac{\overleftarrow\partial}{
\partial X^\mu}
+ \frac{ip_\mu}\hbar) - m_0\} =
&-&S^{R}(X,p)\hat\Lambda \Sigma^{-+}(X,p) \nonumber \\ & +&
S^{-+}(X,p)\hat\Lambda\Sigma^{A}(X,p), \nonumber \\
\end{eqnarray}
\begin{equation}
\hat\Lambda = \exp\left(-\frac{i\hbar}2(
\frac{\overleftarrow\partial}{\partial X^\mu}\frac
{\overrightarrow\partial}{\partial p_\mu} - \frac{
\overleftarrow\partial}{\partial p_\mu} \frac{\overrightarrow\partial}
{\partial X^\mu})\right).
\label{hatlambda}
\end{equation}
 Now subtracting and adding these resulting equations, one arrives at two
futher equations, which we identify as the transport and constraint
equations
respectively:
\begin{equation}
 \frac{i\hbar}2\{\gamma^\mu,\frac{\partial
S^{-+}}{\partial X^\mu}\} + [\not p,S^{-+}(X,p)] =
I_{-}
\label{full1}
\end{equation}
and  
\begin{equation}
 \frac{i\hbar}2[\gamma^\mu,\frac{ \partial S^{-+}}{\partial
X^\mu}] + \{ \not p - m_0,S^{-+}\} = I_{+}.
\label{full2}
\end{equation}
In these equations, the terms that occur on the right hand side are
decomposed into three types of contribution, one containing at least
one retarded function, one with at least one advanced function and a 
further term with neither, which in the semi-classical limit is the
origin of the collision integral.   Explicitly, one has
\begin{equation}
I_{\mp} = I_{{\rm coll}} + I^A_{\mp} + I^R_{\mp},
\label{is}
\end{equation}
with
\begin{eqnarray}
I_{{\rm coll}} &=&
\Sigma^{-+}(X,p) \hat\Lambda S^{+-}(X,p) - \Sigma^{+-}(X,p) \hat\Lambda
S^{-+}(X,p) \nonumber \\
&=& I_{{\rm coll}}^{{\rm gain}} - I_{{\rm coll}}^{{\rm loss}},
\label{collision}
\end{eqnarray}
\begin{equation}
I^R_{\mp} =  - \Sigma^{-+} ( X,p ) \hat \Lambda
                                  S^{\rm R} ( X,p )
         \pm S^{\rm R} ( X,p ) \hat\Lambda
                                         \Sigma^{-+} ( X,p )
\label{irmp}
\end{equation}
and
\begin{equation}
I^A_{\mp} =\Sigma^{\rm A} ( X,p ) \hat\Lambda
                                  S^{-+} ( X,p )
         \mp S^{-+} ( X,p ) \hat\Lambda
                                      \Sigma^{\rm A} ( X,p ).
\label{iamp}
\end{equation}
Equations (\ref{full1}) and (\ref{full2}) are the central, exact equations
that describe the non-equilibrium evolution of a system
of interacting quarks.   To actually see that
these are in fact transport and constrint equations known from Vlasov
of Boltmann theory requires some (hard) work.   This follows only under
certain approximations, and of course one needs some model in order to
specify the interactions.
For this purpose, we will use the Nambu--Jona-Lasinio
model.   Before doing this however, note that an exact solution 
of Eqs.(\ref{full1}) and (\ref{full2})
follows
formally on making a spinor decomposition,
\begin{equation}
-i\hbar S^{-+} = F + i\gamma_5 P + \gamma^\mu V_\mu + \gamma^\mu\gamma_5
A_\mu =\frac
12\sigma^{\mu\nu}S_{\mu\nu}.
\label{dec} 
\end{equation}
The equations for the projected functions $F\sim {\rm tr} S^{-+}$, 
$P\sim {\rm tr} \gamma_5 S^{-+}$, \dots, form a set of 16 times 2 coupled
equations that need to be solved simultaneously.   This is not only a 
formidable task from the computational point of view, it also offers at
present little physical insight.

For reasons of simplicity, therefore, we introduce the quasiparticle
ansatz that contains the quark and antiquark distribution functions
$f_q(X,p)$ and $f_{\bar q}(X,p)$, and which puts these on their mass
shell,
\begin{equation}
S^{-+}(X,p) = 2\pi i\frac {\not p + m}{2E_p}[\delta(p_0 - E_p)f_q(X,p) -
 \delta(p_0 + E_p)\bar f_{\bar q}(X,-p)]
\label{smp}   
\end{equation}
with $\bar f_{q,\bar q} = 1-f_{q,\bar q}$.   Similar expressions can also be 
easily written down for the remaining components of the matrix $S^{ij}$.

\subsection{The Vlasov equation for the NJL model}

At this point, one cannot go futher unless one specifies a theory or model
from which the self-energy can be calculated.   A four point interaction
like that of the SU(2) NJL model is particularly simple to handle because
the
Feynman rules are particularly simple: (a) a directed line represents a
fermion.   The signs attributed to the beginning ($i$) and end $(j)$
of the line reflect in the Green function $iS^{ji}$ to be associated with 
the line.   (b) an interaction line can have only a single sign on both
of its ends.
If the sign is $\pm$, it is to be translated as $\pm iV$, with $V$ being the
interaction strength.   In the NJL model, this is $V=-2G$.

According to these rules, in the Hartree approximation, it follows 
immediately that  $\Sigma^{+-}=
\Sigma^{-+} = 0$, so that $I_{\rm coll} = I_R = 0$.   Only $I_A\ne0$.
Furthermore $\Sigma^A(X,p) = \Sigma^A(X) = m(X)$ alone, so that
\begin{eqnarray}
I_-^A &=& \Sigma^A(X)[1-\frac {i\hbar}2(\overleftarrow\partial_x
\overrightarrow \partial_p - \overleftarrow\partial_p\overrightarrow
\partial_x)]S^{-+} \nonumber \\
&-& S^{-+}[1-\frac {i\hbar}2(\overleftarrow \partial_X\overrightarrow\partial_
p - \overleftarrow\partial_p\overrightarrow\partial_X)]\Sigma^A(X) +
O(\hbar^2)
\nonumber \\
\end{eqnarray}
or
\begin{equation}
I_-^A= -i\hbar\partial_\mu\Sigma^A\partial_p^\mu S^{-+},
\end{equation}
and the transport equation becomes 
\begin{equation}
\frac{i\hbar}2\{\gamma^\mu\frac{\partial S^{-+}(X,p)}{\partial X^\mu}\}
+ [\not\! p, S^{-+}(X,p)] = -i\hbar\partial_\mu\Sigma^A\partial_p^\mu
S^{-+}.
\label{e:vlas}
\end{equation}
Assuming that the quasiparticle ansatz for $S^{-+}(X,p)$ of Eq.(\ref{smp})
holds and that the mass is to be considered as the dynamically generated
Hartree mass that is to be self-consistently determined, one can insert
Eq.(\ref{smp}) into Eq.(\ref{e:vlas}), take the trace over spinor indices
and integrate over a positive energy interval $\Delta_+$ that contains
$E_p$, to arrive at an equation for the quark distribution function,
\begin{equation}
\frac{\partial}{\partial X^0} f_q(X,\vec p) = p_i\partial^i\frac {f(X,\vec p)}
{E_p} + m(X) \partial_i m(X) \partial_p^i(\frac{f_q(X,p)}{E_p}) =0.
\end{equation}
On performing the derivatives and extracting a factor of $1/E_p$, one
can write this as
\begin{equation}
\frac 1{E_p}(p^\mu\partial_\mu f_q(X,\vec p) + m(X)\partial_\mu
m(X)\partial_p^\mu f_q(X,\vec p)) = 0,
\label{e:vlasov}
\end{equation}
which is the Vlasov equation for the model.   It must be solved concurrently
with the gap equation for $m(X)$,
\begin{equation}
m(X) = m_0 + 4GN_cm(X)\int\frac{d^3p}{(2\pi\hbar)^3} \frac 1{E_p(X)} [ 1 -
f_q(X,p)
- f_{\bar q}(X,p)],
\label{gap}
\end{equation}
that is derived directly from the Hartree self-energy.    The constraint
equation, in this same approximation in the expansion in $\hbar$, is
\begin{equation}
(p^2-m^2(X))f_q(X,p) = 0,
\label{e:constr}
\end{equation}
which validates our use of the quasiparticle assumption as being exact. 
Equation (\ref{e:vlasov}) indiates that chiral symmetry breaking enters
via the condensate or mass already as a spatially varying potential in the
Vlasov equation.

\subsection{The Boltzmann equation for the NJL model}

In principle, the next step from a physical point of view would be to 
incorporate all self-energy diagrams of the next order in $1/N_c$.   This 
would correspond to meson exchange \cite{quack}.    This has not been done
yet formally \cite{leshouches} and we will touch on this briefly in the
following subsection.    Here we shall rather examine the simpler problem of
considering our self-energy with at least two interaction vertices, such as
shown in Figs.~17 and 18 for the NJL model.    These are the minimal
types of diagram that can possibly give rise to an off-diagonal self-energy
$\Sigma^{+-}$ say, and therefore to a non-vanishing contribution to the gain 
and loss terms that comprise $I_{\rm coll}$ in the transport equation,
Eq.(\ref{full1}).   We will not give details here, but just note the
salient features \cite{ogu}.

%
%
\begin{center}
\begin{figure}
\epsfig{file=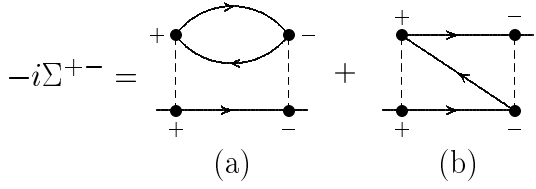,angle=0,scale=0.5}
\caption{Direct (a) and exchange (b) graphs that contribute to
$\Sigma^{-+}$ and which contain two interaction lines.   The vertices
can be either all scalar or all pseudoscalar in nature.}
\end{figure}
\end{center}
 
%
%
\begin{center}
\begin{figure}
\epsfig{file=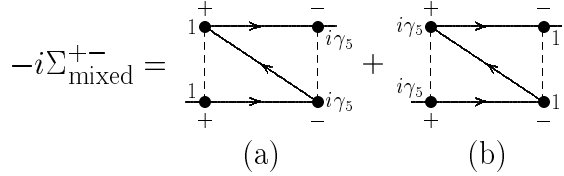,angle=0,scale=0.5}
\caption{Mixed graphs that contribute to $\Sigma^{+-}_{{\rm mixed}}$ and 
which contain two interaction lines.}
\end{figure}
\end{center}

Firstly, a direct translation of the off-diagonal graphs, in the scalar
channel say,
\begin{eqnarray}
\Sigma_\sigma^{+-}(X,p) &=& - 4G^2\hbar^2\int \frac{d^4p_1}{(2\pi\hbar)^4}
\frac{d^4p_2}{
(2\pi\hbar)^4}\frac{d^4p_3}{(2\pi\hbar)^4}
(2\pi\hbar)^4\delta(p-p_1+p_2-p_3) \nonumber \\
      &\times& [ S^{+-}(X,p_1) {\rm tr}\left( S^{-+}(X,p_2) S^{+-}(X,p_3)
\right)
\nonumber \\
& &-S^{+-}(X,p_1)S^{-+}(X,p_2)S^{+-}(X,p_3)  ],
\label{sigsig}
\end{eqnarray} 
contains a product of three Green functions.   Recalling that this will be 
multiplied by $S^{-+}$ in the collision integral and also that we must trace
and integrate the result first over a positive energy interval,
we can easily see that such a procedure
 will lead
to eight terms that each contain some product of four quark or antiquark
distribution functions, such as for example
\begin{equation}
{\rm coefficient} \times \bar f_q(p) f_q(p_2) \bar f_q(p_3) f_q(p_4)
\label{e:stuff}
\end{equation}
In a loose sense, if one designates $f_{q,\bar q}(p)$ to represent an
incoming quark (antiquark) and $\bar f_{q\bar q}$ to represent an outgoing
quark (antiquark), then one can draw diagrams associated with each process.
For example, the products listed in Eq.(\ref{e:stuff}) would represent
quark-quark scattering.   A similar term of the eight possible 
leads to quark-antiquark
scattering,
while the remaining six that are not listed (but which are easily worked
out), are shown in Fig.~19.   Some of these look like the
 typical vacuum fluctuation
processes that would occur in any relativistic theory and in addition to
these, there are others that  give
rise to pair creation and annihilation.    All six graphs of this figure can
be shown to vanish from energy-momentum conservation due to the
quasiparticle assumption!    This gives us an indication of the complexity
and richness of the theory that would go beyond the standard collision 
scenario if one relaxes this assumption.

%
%
\begin{center}
\begin{figure}
\epsfig{file=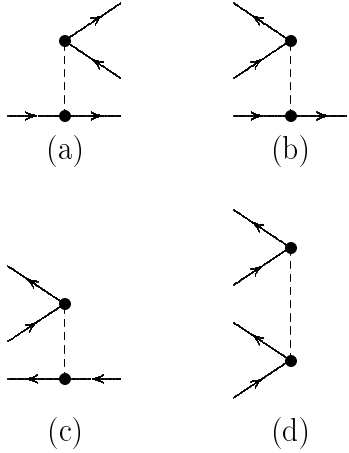,angle=0,scale=0.5}
\caption{Six graphs that arize from the term $\sim \Sigma^{+-}S^{-+}$.
These are heuristic graphs are are not Feynman diagrams.}
\end{figure}
\end{center}

Secondly, it is important to verify that the coefficient functions in the
term of (\ref{e:stuff}) in fact truly give rise to the differential 
cross section for elastic quark-quark 
scattering as would be calculated from real
Feynman diagrams (and not heuristic graphs of Fig.~16) such as are displayed
in Fig.~20.   In fact, this has been explicitly demonstrated to be the case
\cite{ogu}.   One finds that the contribution from Fig.~17(a) gives rise to
the amplitudes squared of both the $s$ or $u$ channels for $q q$
scattering (or $s$ or $t$ channels for $q\bar q$ scattering), while Fig.~17(b)
is required to produce the interference terms between them.
It appears that evaluating nonequilibrium self-energies for the Boltzmann
equation leads to scatering processes that can be obtained from all possible
combinations of cutting the slef-energy grphas of Fig.~17 vertically, 
reminiscent of the Wick-Cutkowsky rules \cite{itzyk}.

%
%
\begin{center}
\begin{figure}
\epsfig{file=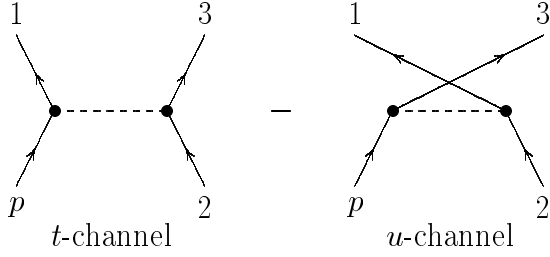,angle=0,scale=0.5}
\caption{$t$ and $u$ channel 
Feynman graphs for elastic quark-quark scattering, to lowest
order in $1/N_c$.}
\end{figure}
\end{center}
 
Finally, one arrives at a Boltzmann equation from Eq.(\ref{full1}).
It reads
\begin{eqnarray}
&&p^\mu\partial_\mu f_q(X,\vec p) + m(X)\partial_\mu m(X)\partial_p^\mu
f_q(X,
\vec p) =\nonumber \\
&&\quad 
N_c\int d\Omega \int \frac {d^3p_2}{(2\pi\hbar)^3 2E_{p_2}} |\vec v_p-\vec
v_2|
2E_p 2E_{p_2}
\nonumber \\
&&\times\{ \frac 12
\frac{d\sigma}{d\Omega}|_{qq\rightarrow qq}(p2\rightarrow 13)(f_q(p_1)\bar
f_q(
(p_3) \bar f_q(p) - \bar f(p_1) f_q(p_2) \bar f_q(p_3) f_q(p)) \nonumber \\
& & +\frac{d\sigma}{d\Omega}|_{q\bar q\rightarrow q\bar q}(p2\rightarrow 13)
(f_q(p_1) \bar
f_{\bar q}(p_2) f_{\bar q}(p_3) \bar f_q(p)
- \bar
f_q(p_1) \bar f_{\bar q}(p_3) f_{\bar q}(p_2) f_q(p))\}, \nonumber \\
\label{boltmann}
\end{eqnarray}
The constraint derived earlier, Eq.(\ref{e:constr}), however, remains unaltered.
From the Boltzmann equation, it is apparent that the changes in the
condensate with the medium affect the equation in two possible places:
(a) As with the Vlasov equation, a medium dependent potential occurs on
the left hand side that is related to the effective quark mass in medium and
(b) the cross-sections occurring on the right hand side are medium
dependent, and also depend on changes of the quark and meson masses in the 
medium.   As we have seen in the preceding section, the cross section for
quark-antiquark scattering diverges at the phase transition.   

The actual answer as to what one should expect from numerical simulations is
however unclear:   since the differential cross-sections are averaged over,
one
may lose the sharp signal of the divergence.   Howver, the force
term
on the left hand side may still play an essential role.    At this stage also,
too many physical features are still lacking, in particular, the coupling
of the quark degrees of freedom to mesons and their coupling back to the
quarks.   This must lead to a hadronization scenario.  In the final 
subsection of this chapter, we briefly sketch how this might occur.
For numerical simulations thus far, we refer the reader to \cite{prjh}
and other references cited therein.

\subsection{Higher orders in $1/N_c$ and meson production}

As already pointed out earlier, the expansion in the coupling strength
that was used for selecting the diagrams of the last section is 
inadmissable, because $G\Lambda\sim 2$.    Going to higher orders in the
$1/N_c$ expansion is however non-trivial, as a {\it symmetry
conserving}\ set of graphs must be chosen.   From \cite{quack,lemmer},
we know that this comprises firstly the set of graphs of 
Fig.~21 for the self-energy, where the ``F'' denotes the new
full Green function that must be newly determined in a self
consistent fashion.     

%
%
\begin{center}
\begin{figure}
\epsfig{file=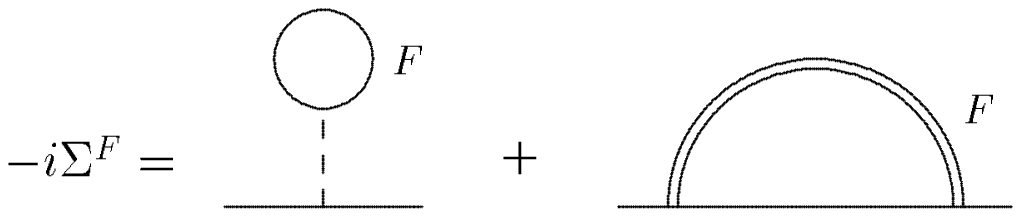,angle=0}
\caption{Self-consistent self-energy that includes meson exchange.}
\end{figure}
\end{center}
 
Denoting the two terms in the self-energy as $\Sigma_1$ and $\Sigma_2$
respectively, one can make an expansion of the new full Green function
about that governed by $\Sigma_1$,
\begin{equation}
\Sigma = \Sigma_1+\Sigma_2(k)
\end{equation}
\begin{eqnarray}
S^F &=& S_{\Sigma_1} + S_{\Sigma_1}\Sigma_2S^F \nonumber \\
&=& \frac 1{\not\! p - \Sigma_1} + \frac 1{\not\! p - \Sigma_1}
\Sigma_2 \frac 1{\not\! p - \Sigma_1} + \dots
\end{eqnarray}
Concomitantly, the irreducible polarization $\Pi^F(k)$ now occurring
in the quark antiquark scattering amplitude 
\begin{equation}
-iD^F(k) = \frac{2iG}{1-2G\Pi^F(k)}
\end{equation}
must contain further terms,
\begin{equation}
\Pi^F = \Pi^0 + \delta \Pi
\end{equation}
where $\Pi^0$ is the simple quark loop, in order to be symmetry 
conserving.   The graphs required for $\Pi^F$ are shown in Fig.~22.

%
%
\begin{center}
\begin{figure}
\epsfig{file=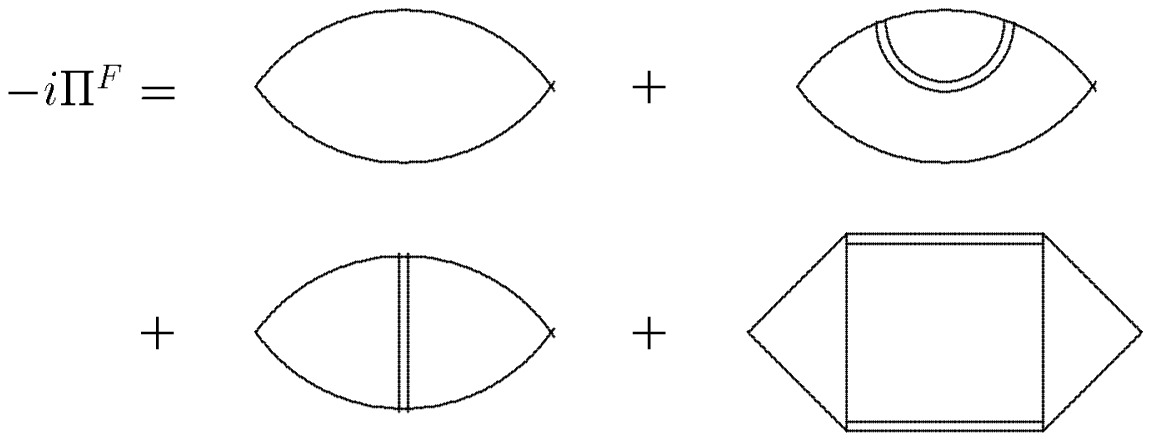,angle=0}
\caption{Contributions to the irreducible polarization to next to
leading order in the $1/N_c$ expansion.}
\end{figure}
\end{center}
 
Inserting the expansion of the Green function and the irreducible
polarization into the full self-energy of Fig.~21, leads to graphs
that contain {\it inter alia} diagrams of the form shown in Fig.~23.

%
%
\begin{center}
\begin{figure}
\epsfig{file=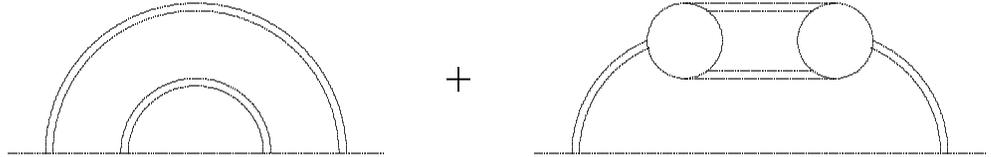,angle=0}
\caption{Diagrams occurring in $\Sigma$, that lead to the hadronization
of a quark into two mesons.}  
\end{figure}  
\end{center}

This gives us an intuitive understanding that, on evaluating these diagrams
in the non-equilibrium scenario, we should no longer simply obtain a 
cross-section for elastic quark-quark and quark-antiquark scattering,
but also the hadronization process of $q\bar q\rightarrow MM'$, where
$M$ and $M'$ are mesons.   Much work however, remains to be done in 
this regard.

\section{Concluding comments}

In this series of lectures, we have investigated some aspects of chiral
symmetry breaking at finite temperatures.    We have seen that in the last
few years, much information is emerging from the lattice gauge community
that tells us about the transition region itself.  
Chiral perturbation theory, on the other hand, while being excellent in the
low temperature regime, cannot adequately describe a phase transition.

In the following section, we have investigated the Nambu--Jona-Lasinio
model at finite temperatures.   It gives a remarkably good qualitative
agreement with the lattice data in the realm of static properties.   It
fails, however, to describe the bulk thermodynamic properties well,
primarily due to the fact that confinement is lacking.   The NJL model gives
a simple picture for a delocalization rather than a deconfinement
transition.
Associated with this (physically appealing) picture that bound mesons become
delocalized at the transition temperature -- now the Mott temperature --
and are still correlated states with a finite width in the quark medium,
are marked divergences in many functions, such as the pion radius,
$\pi$-$\pi$ and $\pi$-$K$ scattering lengths (not discussed here), as well
as the phenomenon of critical scattering, observed in the quark-antiquark
channel.

Due to the fact that none of the apparent singularities are directly
observable experimentally, we have turned to transport theory, in order to
investigate what effects are to be expected from a condensate density that
is medium dependent.   Calculations at this stage indicate that a Boltzmann
equation is dependent on the condensate through a force term, and also via
the cross-sections that arize from binary collisions among the quarks and
antiquarks.   Howver, the stage of calculation is still primitive: a
consistent physical theory that includes mesons  and which overcomes the
problems associated with the lack of confinement is required before one can
expect to obtain credible results.   This, of course, leaves the path open
for future research.

\section{Acknowledgments}

I would like to thank Jean Cleymans for the opportunity of being able to
speak in Cape Town and for providing a comfortable and stimulating 
scientific atmosphere.     In preparing this manuscript, I am indebted to
both E. Laermann and P. Rehberg for providing some of the figures in 
postscript form.   A hearty thanks also goes to G. Papp for his concerted 
efforts and timely thinking in nursing a collapsing computer.
This work has been supported in part
  by the Deutsche Forschungsgemeinschaft
DFG under the contract number Hu 233/4-4, and by the German Ministry for
Education and Research (BMBF) under contract number 06 HD 742.

\end{document}